 \documentclass[preprint2,longabstract]{aastex}

\slugcomment{Submitted to the Astronomical Journal}

\shorttitle{Keel et al.}
\shortauthors{Faded Quasar: Hanny's Voorwerp with HST}

\begin{document}


\title{The History and Environment of a Faded Quasar: {\it Hubble Space Telescope} observations of Hanny's Voorwerp and IC 2497\footnote{Based on observations with the NASA/ESA {\it Hubble Space Telescope}
obtained at the Space Telescope Science Institute, which is operated
by the Association of Universities for Research in Astronomy, Inc.,
under NASA contract No. NAS5-26555.}}


\author{William C. Keel\altaffilmark{1}}
\affil{Department of Physics and Astronomy, University of Alabama, Tuscaloosa, AL 35487}
\email{wkeel@ua.edu}

\author{Chris J. Lintott}
\affil{Astrophysics, Oxford University, Denys Wilkinson Building, Keble Road, Oxford, OX1 3RH, UK}

\author{Kevin Schawinski\altaffilmark{2}}
\affil{Department of Physics, Yale University}

\author{Vardha N. Bennert\altaffilmark{3}}
\affil{Department of Physics, University of California, Santa Barbara}

\author{Daniel Thomas}
\affil{University of Portsmouth}

\author{Anna Manning\altaffilmark{1}}
\affil{Department of Physics and Astronomy, University of Alabama, Tuscaloosa, AL 35487}

\author{S. Drew Chojnowski\altaffilmark{1,4,5}}
\affil{Department of Astronomy,  University of Virginia}

\author{Hanny van Arkel}
\affil{Citaverde College, Heerlen, The Netherlands}

\and

\author{Stuart Lynn\altaffilmark{1}}
\affil{Adler Planetarium, 1300 South Lake Shore Drive, Chicago, IL 60605}


\altaffiltext{1}{Visiting Astronomer, Kitt Peak National Observatory, National Optical Astronomy Observatories, which is operated by the Association of Universities for Research in Astronomy, Inc. (AURA) under cooperative agreement with the National Science Foundation.
The WIYN Observatory is a joint facility of the University of Wisconsin-Madison, Indiana University, Yale University, and the National Optical Astronomy Observatory.}

\altaffiltext{2}{NASA Einstein Fellow}

\altaffiltext{3}{Current address: Physics Department, California Polytechnic State University, San Luis Obispo, CA 93407.}

\altaffiltext{4}{Participant in SARA Research Experiences for Undergraduates program, funded by the National Science Foundation.}

\altaffiltext{5}{Current address: Dept. of Astronomy, University of Virginia, Charlottesville, VA 22904.}


\clearpage

\begin{abstract}
We present {\it Hubble Space Telescope} imaging and spectroscopy, along with supporting GALEX
and ground-based data, for the extended high-ionization cloud known as Hanny's Voorwerp, near
the spiral galaxy IC 2497. WFC3 images show complex dust absorption near the nucleus of IC 2497.
The galaxy core in these data is, within the errors, coincident with the VLBI core component
marking the active nucleus. STIS optical spectra show the AGN to be a type 2 Seyfert
of rather low luminosity. The derived ionization parameter log $U = -3.5$ is in
accord with the weak X-ray emission from the AGN. We find no high-ionization gas near the nucleus, adding to the evidence that
the AGN is currently at a low radiative output (perhaps with the central black hole having switched to a mode dominated
by kinetic energy). The nucleus is accompanied by an expanding 
ring of ionized gas $\approx 500$ pc in projected diameter on the side opposite Hanny's Voorwerp. Where sampled by the STIS slit, this ring has 
Doppler offset $\approx 300$ km s$^{-1}$ from the nucleus, implying a kinematic age
$< 7 \times 10^5$ years.  Narrowband [O III] and H$\alpha$+[N II] ACS images show fine structure
in Hanny's Voorwerp, including limb-brightened sections suggesting modest interaction with a galactic outflow 
and small areas where H$\alpha$ is strong. We identify these latter regions as
regions ionized by recent star formation, in contrast to the AGN ionization of the entire
cloud. These candidate ``normal" H II regions contain blue continuum objects, whose
colors are consistent with young stellar populations; they appear only in a 2-kpc region toward 
IC 2497 in projection, perhaps meaning that the star formation was triggered by compression from a 
narrow outflow.
The ionization-sensitive ratio [O III]/H$\alpha$ shows broad bands across the object at a skew
angle to the galaxy nucleus, and no discernible pattern near the prominent ``hole" in the ionized gas. 
The independence of ionization and surface brightness suggests that there is substantial spatial structure
which remains unresolved, to such an extent that the surface brightness samples the number of
denser filaments rather than the characteristic density in emission regions; this might be a typical
feature of gas in tidal tails, currently measurable only when such gas is highly ionized.
These results fit with our picture of an ionization echo from an AGN whose ionizing luminosity has
dropped by a factor $>100$ (and possibly much more) within the last $1-2 \times 10^5$ years; we suggest 
a tentative sequence of events
in IC 2497 and discuss implications of such rapid fluctuations in luminosity
for our understanding of AGN demographics.
\end{abstract}


\section{Introduction}

The central energy sources of active galactic nuclei (AGN) are known to vary on a wide
range of timescales. Direct observation samples variations  from hours to decades (sometimes strong, and including 
dramatic changes in the prominence of the broad-line region). The dramatic
evolution of luminous AGN with redshift demonstrates cosmological evolution, involving the
entire AGN population. What this evolution entails for individual AGN is only indirectly
constrained. Several arguments suggest that the central black holes grow
episodically; the duty cycles, amount of accretion, or timescales for these episodes
decrease with cosmic time (\citealt{Martini2004}, \citealt{Hopkins2005}). Luminous QSOs cannot maintain the observed energy output
for much of cosmic history without the black holes becoming more massive than any we
observe, and evidence for an excess of interaction signatures in QSO host galaxies (although
not clearly present at lower luminosities)
suggests that accretion is enhanced over roughly the timespan that we can recognize these 
signatures (typically a few times $10^8$ years). However, none of these factors reveals how the
energy output behaves over timescales of $10^3$--$10^7$ years, spanning values for
activity scales in quasars suggested by some observational considerations  (\citealt{Martini2003}, \citealt{Kirkman}) as well as
calculations of instabilities in accretion disks (\citealt{Shields1978}, \citealt{Goodman},
\citealt{Janiuk}, \citealt{Done}). Timescales of accretion-disk behavior may be estimated from scaling
the state changes seen in X-ray binaries containing stellar-mass black holes (\citealt{Maccarone},
\citealt{McHardy}.

We describe here new observations of an object which may hold key insights to otherwise
unprobed timescales in the history of individual AGN - Hanny's Voorwerp.
Among the signature serendipitous discoveries of the Galaxy Zoo project \citep{Lintott08}, this is a giant
high-ionization nebula near the bright spiral galaxy IC 2497 \citep{Lintott09}. It was reported on the project
forum\footnote{www.galaxyzooforum.org} by citizen scientist and co-author Hanny van Arkel, only a few weeks into the Galaxy Zoo examination of the SDSS
main galaxy sample. Followup observations revealed a unique combination of characteristics,
indicating that this object traces a luminous AGN which must be unusual either in obscuration or history. 

We first briefly summarize results from \cite{Lintott09}, \cite{Josza}, \cite{Rampadarath}, and \cite{Schawinski2010}. Hanny's
Voorwerp is a region of highly-ionized material 18 by 33 kpc in projected extent, extending at least
50 kpc from IC 2497 and closely matching its redshift $z=0.050$.
The electron temperature measured from [O III] lines indicates that it is photoionized rather
than shock-ionized.
Such emission-line ratios as He II/H$\beta$ and [Ne V]/[Ne III] show that the ionizing continuum
is hard like an AGN rather than hot stars, while the ionizing luminosity for a source in IC 2497 must be
of order $2 \times 10^{45}$ erg s$^{-1}$ to give the observed ionization parameter and
intensity of recombination lines. However, the nucleus of IC 2497 has very
modest luminosity, with emission lines implying ionizing luminosity $< 10^{40}$ erg s$^{-1}$.  
H I observations show that Hanny's Voorwerp is a small part of a 300-kpc structure around the
southern side of IC 2497 containing $9 \times 10^9$ M$_{\odot}$ of neutral hydrogen. The nucleus
hosts a compact VLBI radio source of modest power, and an additional feature which could be
the brightest knot in a jet pointing roughly toward Hanny's Voorwerp. Lower-resolution radio 
continuum data show what may be a broad outflow in the same direction. 
The ionizing continuum required is of a luminosity associated with QSOs, making IC 2497
a very nearby QSO host galaxy observable in great detail.

The spectra in \cite{Lintott09} indicate substantially subsolar abundances
in Hanny's Voorwerp (and by extension in the H I tail). For gas 
photoionized by an AGN continuum across the relevant range of ionization parameter $U$, 
the [N II]/H$\alpha$ ratio is almost purely an abundance 
indicator. Models by \cite{Groves06}, \cite{FuStockton07}, and
\cite{FuStockton} using the MAPPINGS code \citep{Groves04} show that this line ratio remains consistent for various ionization levels relevant 
to AGN narrow-line regions and extended regions. From Fig. 11 of
\cite{FuStockton}, Hanny's Voorwerp has N/H $0.3 \pm 0.1$ solar.

These data led to two competing interpretations - that the AGN is either hidden or faded.
\cite{Lintott09} introduced the ionization-echo hypothesis, driven by the lack of detected X-rays 
from IC 2497, lack of any high-ionization gas observed in the galaxy, and the shortfall between the
expected far-infrared luminosity and what is observed if most of the AGN output is absorbed
by dust. \cite{Josza} favored a picture in which an outflow, seen in the radio continuum,
cleared a path for ionizing radiation to escape circumnuclear obscuration, so that the
Voorwerp would be ionized by an extant AGN which is so deeply obscured from our
direction that not even the soft X-rays detectable by {\it Swift} would emerge. Evidence that 
QSOs can fade so rapidly would have a significant impact on our understanding of the demographics of QSOs.
IC 2497 is much nearer than any known QSO of its inferred luminosity, yet has managed to 
elude all standard ways of surveying for them, and it is unlikely that a very rare event
would be represented so close to us. This is likely to represent a phenomenon which is so common
among AGN as to alter our estimates of their overall behavior.

The key role of emerging X-rays in distinguishing between fading and obscuration motivated
a set of {\it XMM-Newton} and {\it Suzaku} observations \citep{Schawinski2010}. The results
support a fading scenario. The nuclear X-ray source in IC 2497 is well fitted by a combination
of hot ISM and an AGN which is essentially unobscured beyond the Galactic H I value of
$1.3 \times 10^{20}$ cm$^{-2}$. There is no detected 6.4-keV K$\alpha$ feature
which characterizes deeply obscured ``reflection" AGN, and the {\it Suzaku} data above 10 keV in 
particular rule out a Compton-thick
AGN luminous enough to ionize Hanny's Voorwerp. The 2-10 keV luminosity of
$4.2 \times 10^{40}$ erg s$^{-1}$ falls short  of the luminosity needed to account for the
ionization of Hanny's Voorwerp, by four orders of magnitude.

As part of our effort to unravel the nature of this system, we have obtained {\it Hubble Space
Telescope} (HST) observations and supporting data, from the mid-ultraviolet to the near-infrared.
These allow us to address the ionization structure of Hanny's Voorwerp, the ionization and kinematics of
gas near the nucleus of IC 2497, and evidence for compression-induced star formation suggesting
that an outflow from IC 2497 is interacting with a portion of Hanny's Voorwerp.

\section{Observations}
We report a variety of HST and additional supporting observations, summarized in Table \ref{tbl-1}. These HST data were associated with program 11620.

\subsection{Continuum imaging - WFC3}

Wide Field Camera 3 (WFC3) images were obtained in three bands selected to minimize the 
contribution of emission lines, in order
to study the structure of IC 2497 and seek evidence for star clusters (of whatever ages) in the Voorwerp or 
elsewhere in the massive H I
tidal stream found by \citet{Josza}. For the UVIS CCD images, two exposures
with 2-pixel dither offsets in each detector coordinate were obtained in each of F225W and F814W. 
After some
experimentation for the best balance between typical signal-to-noise ratio and cosmic-ray rejection, we used a {\tt multidrizzle} product \citep{mdriz}
with 2$\sigma$ rejection for cosmic rays and a 2-pixel growing radius around identified cosmic rays. With only two exposures,
a large number of pixels are still affected by cosmic-ray events; these were patched interactively for display purposes.

The near-IR F160W image used multiple readouts (in the STEP100 sequence) and a three-point dither pattern to yield a well-sampled combined image with excellent 
dynamic range and cosmic-ray rejection.

\subsection{Emission-line imaging - ACS}

We used the tunable ramp filters on the Advanced Camera for Surveys (ACS) to isolate [O III] $\lambda$ 5007 and H$\alpha$ at the system redshift, with
filter half-transmission width nominally 2\% (105 and 138 \AA , respectively). The orientation was constrained so that the inner parts of IC 2497, as well as all parts of Hanny's Voorwerp identified from ground-based imaging, fell within the $40 \times 80$" monochromatic field of view. Ground-based data showed the weakness of the continuum compared to these lines, so no matching continuum data were obtained. Processing these images took special care; cumulative radiation damage to the CCDs has resulted in charge-transfer tails from cosmic-ray events affecting much of the area of each image. The best correction 
for this effect has been presented by \cite{CTE}: a constrained deconvolution process which has the net effect of restoring charge in the near-exponential
tails back to its original pixel. We used the {\tt PixCteCorr} routine within PyRAF to apply this correction on the individual ACS images,
followed by drizzle combination with cosmic-ray rejection. Restoration of the CTE charge to its original pixel made cosmic-ray rejection much
more effective.

As an early check on the role of charge-transfer trails, before the \cite{CTE} routine was released, we 
used a multistep heuristic process, incorporating ground-based images to restore structure at low surface brightness. We used similar drizzle parameters to combine the images. Then a large median filter in a blank region of the images was used to model and remove low-level additive banding appearing in the images. The cosmic-ray trails were removed by subtracting the result of a 31-pixel (1.5") median filter along the axis closely matching detector $y$, applied only to pixels with values lower than 0.01 count/second (roughly $S/N < 4$). This step removed real structures at lower surface brightness. To restore these, we used ground-based images in $V$ (dominated by [O III]) and H$\alpha$ (Lintott et al. 2009) as follows: the ACS images were convolved with Gaussian filters to approximate the ground-based PSFs, and the difference between (scaled) ground-based and ACS data was masked in the bright regions where the filtering did not apply. Then this masked difference was added back to the processed ACS images. Finally, residual cosmic rays were patched interactively. This last process was applied only to the region around Hanny's Voorwerp; the central regions of IC 2497 were much brighter than the cosmic-ray residuals and were not so treated. This processing sequence left faint residual streaks from the cosmic-ray response in the Voorwerp, aligned with the columns of the chip (position angle 112$^\circ$). However, {\tt PixCteCorr} gave superior 
results, with no low-surface-brightness artifacts evident on comparison of the two results, so we present further analysis based on the CTE-corrected data.

The relative calibration of the emission-line images was extracted from the passbands generated by the STSDAS {\tt calcband} task. In energy units, the intensity ratio of the two lines corresponding to equal count rates is the inverse ratio of peak throughputs (since the emission lines were placed at the band peaks) multiplied by the ratio of photon energies. From this calibration, equal count rates in the two lines corresponds to an intensity ratio in energy units [O III]/(H$\alpha$+[N II] )= 0.60.
	
\subsection{Astrometric registration of images}

To improve the registration of the HST images for comparison with the radio reference frame,
we compared the positions of stars retrieved from the WFC3 images with
their SDSS coordinates, rejecting stars whose offset
exceeds  0.1" due to proper motion or unresolved duplicity (2 in the UVIS field, one in the smaller IR field). We find coordinate
corrections (from the HST to SDSS systems) given by
$\Delta \alpha = 0.0147 \pm 00016$$^{\rm s}$,
$\Delta \delta = -0.057 \pm 0.046$" for the F814 image, and
$\Delta \alpha = 0.0299 \pm 00021$$^{\rm s}$,
$\Delta \delta = -0.127 \pm 0.034$" for the F160W image. The absolute
accuracy of the SDSS coordinate frame against the USNO astrometric network is 0.045" \citep{Pier}, 
which is the limiting factor in our comparison against the radio reference frame.

\subsection{STIS spectroscopy}

We obtained moderate-resolution spectra using the Space Telescope Imaging Spectrograph (STIS) with
gratings G430L and G750L. The 0.2" slit (aperture 52X0.2) was used to ease
acquisition tolerances in a complex region, and improve surface-brightness
sensitivity to emission lines around the galaxy nucleus. A 100-second ``white-light" acquisition image was used for slit placement, with the options 
ACQTYPE=DIFFUSE, DIFFUSE-CENTER=FLUX-CENTROID specified 
so as to center the slit even if the central light distribution proved to be asymmetric (as it did). 
The $5 \times 5$-arcsecond acquisition image, in the very broad band redward of 5500 \AA\ provided by the long-pass filter, also proved 
useful in comparison with the narrowband images of the center of IC 2497.
For the red spectrum, a fringe flat was obtained during Earth occultation immediately after the F750L exposures. The slit orientation was along celestial position angle 59.8$^\circ$, dictated by scheduling and power constraints; slit positions including both the nucleus and either the southwestern star-forming regions within IC 2497 or the Voorwerp itself were not feasible. Fortuitously, this orientation did sample an extended emission-line loop near the nucleus (section 3.3).

Each grating had two exposures within an orbital visibility period (again, a compromise
between signal-to-noise and efficiency of cosmic-ray rejection); for display purposes, we interactively patched residual cosmic rays on a single-pixel basis.
Our redshift values incorporate use of vacuum wavelengths in STIS spectra, optical as well as ultraviolet.
	
\subsection{Supporting data}

\subsubsection{Ground-based imagery}

We use ground-based emission-line images in our heuristic check on the loss of low-surface-brightness structures in processing the ACS narrowband images (as detailed above).
The H$\alpha$ image, from the Kitt Peak 2.1m telescope, was shown by Lintott et al. (2009). We add a set of BVI exposures (where the V flux
from the Voorwerp is thoroughly dominated by [O III] emission) obtained using the
OPTIC rapid-guiding camera (based on orthogonal-transfer CCDs;
\citealt{OTCCD}) at the 3.5m WIYN telescope. The images
were obtained in November 2008, near zenith passage (just outside the altazimuth tracking 
``hole" overhead) and had image quality reaching 0.45" FWHM in V, well
sampled by 0.14" pixels. The V image is additionally useful in verifying the lack of significant
[O III] emission outside the field of the ACS images.

\subsubsection{Ground-based spectroscopy}

We use a long-slit spectrum obtained with the Goldcam spectrograph at the KPNO 2.1m
telescope to confirm the association of IC 2497 and its apparent companion galaxy to the E, and the nature
of emission regions to the southwest of the nucleus of IC 2497. A 45-minute exposure was
obtained in June 2010, covering the wavelength range 3400-5600 \AA\ . The dispersion
was 1.25 \AA\ pixel${-1}$, with FWHM resolution 4.1 \AA\ . The slit was oriented in position angle 91$^\circ$, 
passing through  the nucleus of the companion galaxy and the H$\alpha$ emission region 
southwest of the nucleus of IC 2497. The companion nucleus shows absorption in the Balmer and Ca II lines,
but no emission. We derive $z = 0.04977 \pm 0.0020$, and a difference between the companion 
and the part of IC 2497 sampled by our slit $\Delta v = -24 \pm 60$ km s$^{-1}$. The IC 2497 measurement 
matches, well within their errors, the redshift given by the STIS spectra at the nucleus of IC 2497. The matching 
galaxy redshifts strongly suggest physical proximity, so that the spiral companion is likely to be physically 
associated with IC 2497 (though its high degree of symmetry makes it unlikely to have been the donor for 
the H I tail; a suitably strong tidal impulse would have probably left
it much more disturbed even after several crossing times).

Especially on the eastern side of the disk of IC 2497, H$\beta$ is consistently narrower than the
single-component FWHM for [O III] $\lambda 5007$.  Multiple components of line emission may be blended together at our spatial
resolution, so that H$\beta$ is relatively stronger in the narrower-line gas. This would also
be observed if the Balmer emission is more strongly confined to star-forming regions than is
[O III], which is likely the case since the nucleus (just to one side of our slit 
position) has significant [O III] emission. The FWHM of the blended [O II] 3727 
doublet is close to that of H$\beta$.
From the emission region to the southwest of the nucleus, making a conservative correction 
for stellar absorption underlying H$\beta$, we measure [O III] $\lambda 5007$/H$\beta$=1.8, 
[O II] $\lambda 3727$/[O III] $\lambda 5007$ = 0.70. These are consistent with normal H II regions in the galaxy's bar.

\subsubsection{GALEX images and spectra}	
We have analyzed GALEX data obtained as part of the Nearby Galaxy Survey \citep{NGS}, 
with long exposures both in direct and spectroscopic modes (Table \ref{tbl-1}). Because Hanny's Voorwerp is such an extended object, we extracted
its integrated spectrum directly from the two-dimensional dispersed
images, using dispersion relations and effective-area values from
\cite{GALEXdata}. A small error in flux calibration across emission lines is introduced by treating
pixels within the line as having wavelengths appropriate
to their offset from the object center (rather than all these pixels having effectively the same wavelength) but this error falls well within the
Poisson errors of the spectra. The GALEX spectra are combined in
Fig. \ref{fig-galex}. The C IV $\lambda 1548+1550$, He II $\lambda 1640$, and C III] $\lambda 1909$ lines 
are clearly detected; [Ne IV] $\lambda 2425$, sometimes seen in the extended emission of radio galaxies, 
may be present at the $\approx 2 \sigma$ level. Their integrated fluxes are compared to optical lines (from Lintott et al. 2009) in Table \ref{tbl-galex}.

In both the optical and UV ranges, the strength of He II lines is notable. The He II $\lambda 4686$/H$\beta$ ratio is in the upper part of range seen for
extended emission-line regions around QSOs \cite{FuStockton}, perhaps suggesting that the ionizing continuum from IC 2497 is at the hard end of the 
range seen for QSOs. The He II $\lambda 1640$/$\lambda 4686$ ratio in AGN can be affected by
resonant effects, since H Ly$\alpha$ nearly coincides in energy with the excitation level of $n=4$ in He$^+$ \citep{MacAlpine}. Such effects
should be much reduced in the low-density, kinematically quiet environment here; the energy difference between H and He energy
levels translates to the Doppler shift from motions at 120 km s$^{-1}$, much larger than local changes seen in our long-slit spectra.
This makes the low-density value for the $\lambda 1640$/$\lambda 4686$  appropriate for the intrinsic line ratio, so we can set a limit to
the reddening. When resonant effects are negligible, the line ratio is 6.8 in energy units (using values from \citealt{AGNAGN}). The
estimated error on the integrated optical line flux from \cite{Lintott09} is 12\%, while \cite{GALEXdata} quote a flux accuracy of 3\% for
spectral regions near the center of the field (as applied here); photon statistics are  a larger contribution in this case,
with S/N=12 for the line. The observed line ratio is then $7.46 \pm 1.1$, implying a reddening limit
$E_{4686-1640} < 0.25$ magnitudes. For a typical Galactic extinction law, this would be $A_V < 0.18$; foreground Galactic extinction
in this direction would be only $A_V =0.04$ following \cite{Schlegel}. This strongly limits any role for imbedded dust in the emission-line regions,
as might be expected at the low densities $n_e < 50$ cm$^{-3}$ found from {S II] emission-line ratios.

\begin{figure*}
\includegraphics[scale=0.6,angle=270]{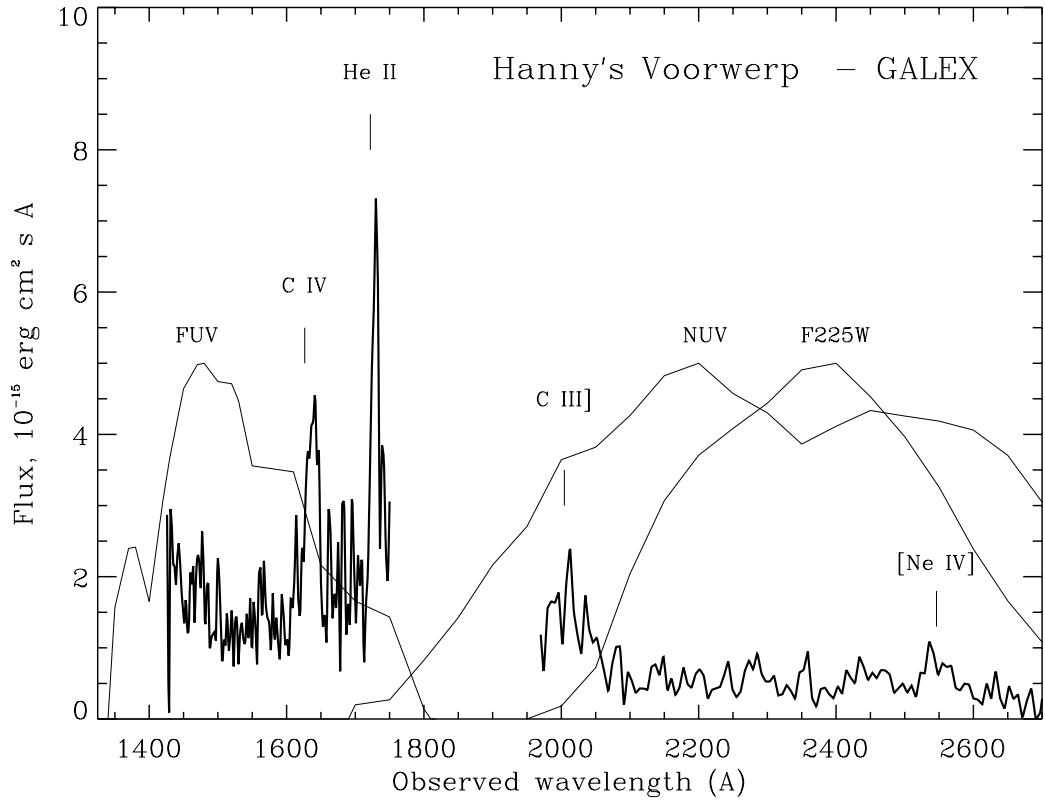}
\caption{Integrated GALEX spectrum of Hanny's Voorwerp from both NUV and FUV grisms, summed over an 18" region N--S. Emission features are marked with the expected center wavelength for $z=0.0498$. Responses of the GALEX imaging filters (FUV, NUV) and the WFC3 F225W filter are superimposed to show how much each is dominated by the continuum.}
 \label{fig-galex}
\end{figure*}

To quantify the roles of line and continuum contributions, we
use the GALEX spectrum to estimate the fractional contribution of emission
lines in each UV filter. We did so by converting the GALEX spectrum to photon
units, multiplying by each filter's response curve, and measuring the difference in total
count rate when emission lines are removed by interpolation. This yields emission-line
fractions of 0.14 in the GALEX FUV image, 0.15 in GALEX NUV, and 0.06 in the WFC3 F225W
filter. In each case, the UV emission is dominated by continuum processes, and strongly
so for F225W. These data confirm that the WFC3 F225W image samples mainly continuum,
as intended. This continuum might have contributions from recombination 
processes, an embedded population of hot stars, or scattered AGN radiation.

\section{Weak nuclear activity in IC 2497}

\subsection {Nuclear location and obscuration}

Given the dust lanes which are prominent on the southeastern side of the galaxy disk, we want to know how much optical obscuration there is toward the nucleus of IC 2497. On large scales, we can address this via comparison of the optical and near-IR HST images, as well as through the astrometric registration of the two VLBI sources reported by Rampadarath et al. (2010).

The nucleus (as defined by peak intensity) at F160W is found within 0.04" of VLBI component C2, with an external error of 0.05" from the scatter in stellar coordinates and systematic error limits on the SDSS frame. This suggests, first, that dust
lanes do not provide strong obscuration toward the nucleus itself at this wavelength, and second, that VLBI component C2 rather than C1 (0.25" away) represents 
the AGN in this galaxy. This makes sense given that C2 has a flatter
spectrum from 6--18 cm, and is surrounded by diffuse emission
roughly aligned with the galaxy disk. As noted by \citet{Rampadarath}, C1 could be a compact knot in a jet directed toward P.A. 215$^\circ$, roughly aligned with both 
the kpc-scale smooth extension in the radio emission (J{\'o}zsa et al. 2009) and the
Voorwerp. 

Obscuration arising close to the nucleus (in a traditional
``torus") would apply to a large solid angle and be accompanied by strong far-IR reradiation, allowing us to construct an energy budget; in contrast, obscuration by a foreground cloud, possibly far from the nucleus, need
not have such an FIR signature. Such ``accidental" obscuration may occur in some type 2 Seyferts, as suggested by the dust morphologies shown by \citet{Malkan}.
This analysis suggests that large-scale foreground obscuring structures are not dense enough to hide an AGN powerful enough to ionize Hanny's Voorwerp.

The F160W near-IR structure shows that the extinction in these dust
lanes is modest at least at this wavelength, so that we have a view to the nucleus which is relatively 
unhindered by spatially-resolved absorbing features. The F814W and F160W images are compared in Fig. \ref{fig-nuc}, resampled to the pixel scale of the F814W data, with a color map  based on convolving the F814W image with the nearest Gaussian equivalent to match the point-spread functions (FWHM 6.1 pixels or 0.23 arcsecond). The flux ratio, in particular, traces intricate
filaments of dust near the nucleus. At the F160W resolution, the deep dust lane near the
nucleus passes close enough to leave the amount of extinction in F814W ambiguous, especially
when including the potential 0.06" positioning error with respect to the core radio source C2
(Fig. \ref{fig-nucoverlay}). When
smoothed to the lower resolution, the implied extinction at F814W ranges from 0.8--1.0 magnitude
across the error  circle of the VLBI core position.
The centroid derived from elliptical isophotes at F160W, at radii up to 0.8 arcseconds, has
offsets in the range $0.06 \pm 0.06$ arcsecond from C2.

\begin{figure*}
\includegraphics[scale=0.7,angle=0]{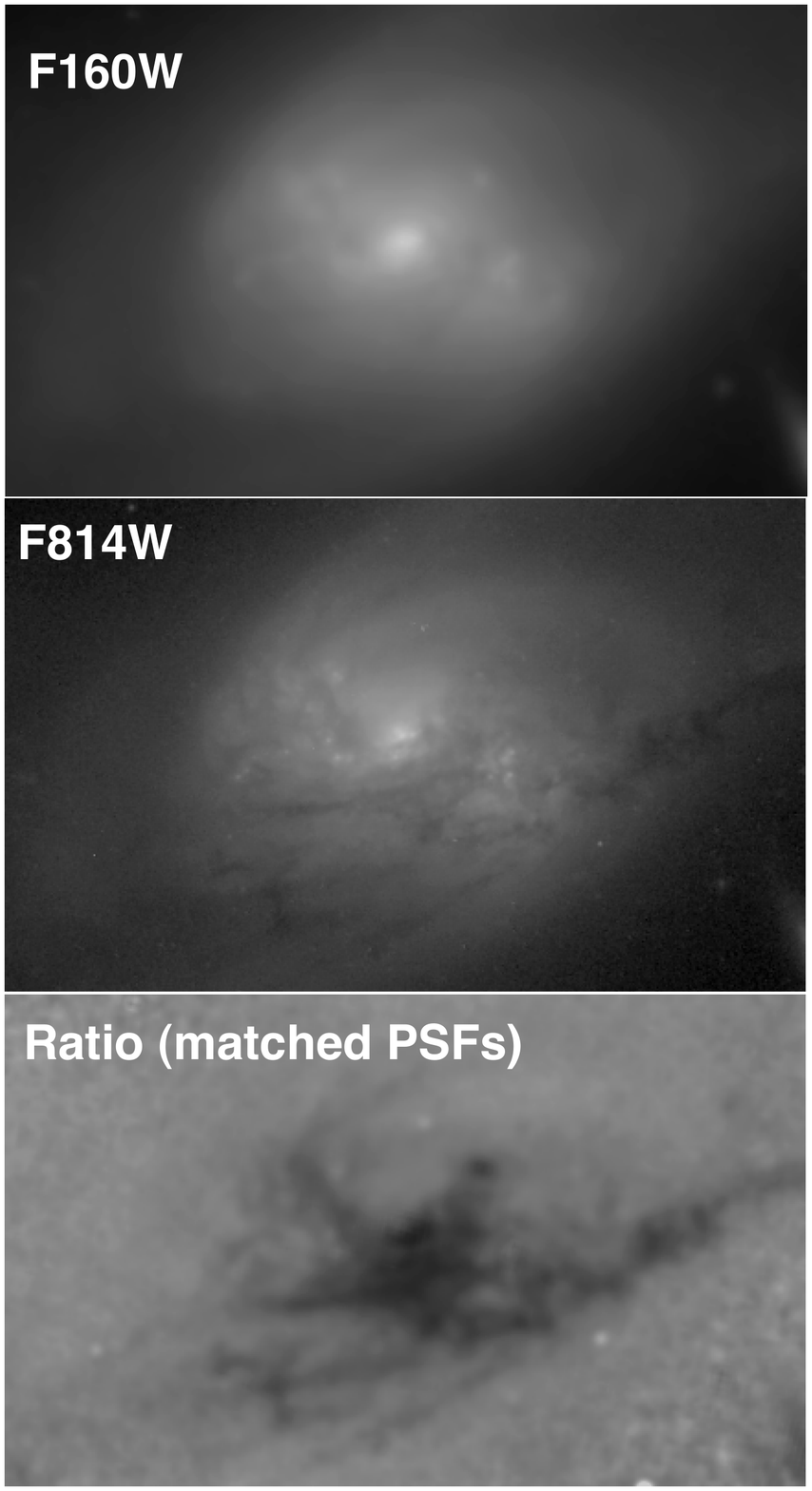}
\caption{The nuclear region of IC 2497 from WFC3 images in F160W and F814W bands shown with logarithmic intensity scales to emphasize structures near the core.
The region shown spans $16.6 \times 10.2$ arcseconds, with north at the top. The bottom panel
traces dust extinction through the flux ratio between these filters on a linear intensity scale, after convolving the F814W image with a Gaussian of 0.23" FWHM to closely match the PSFs.}
 \label{fig-nuc}
\end{figure*}

\begin{figure*}
\includegraphics[scale=0.6,angle=0]{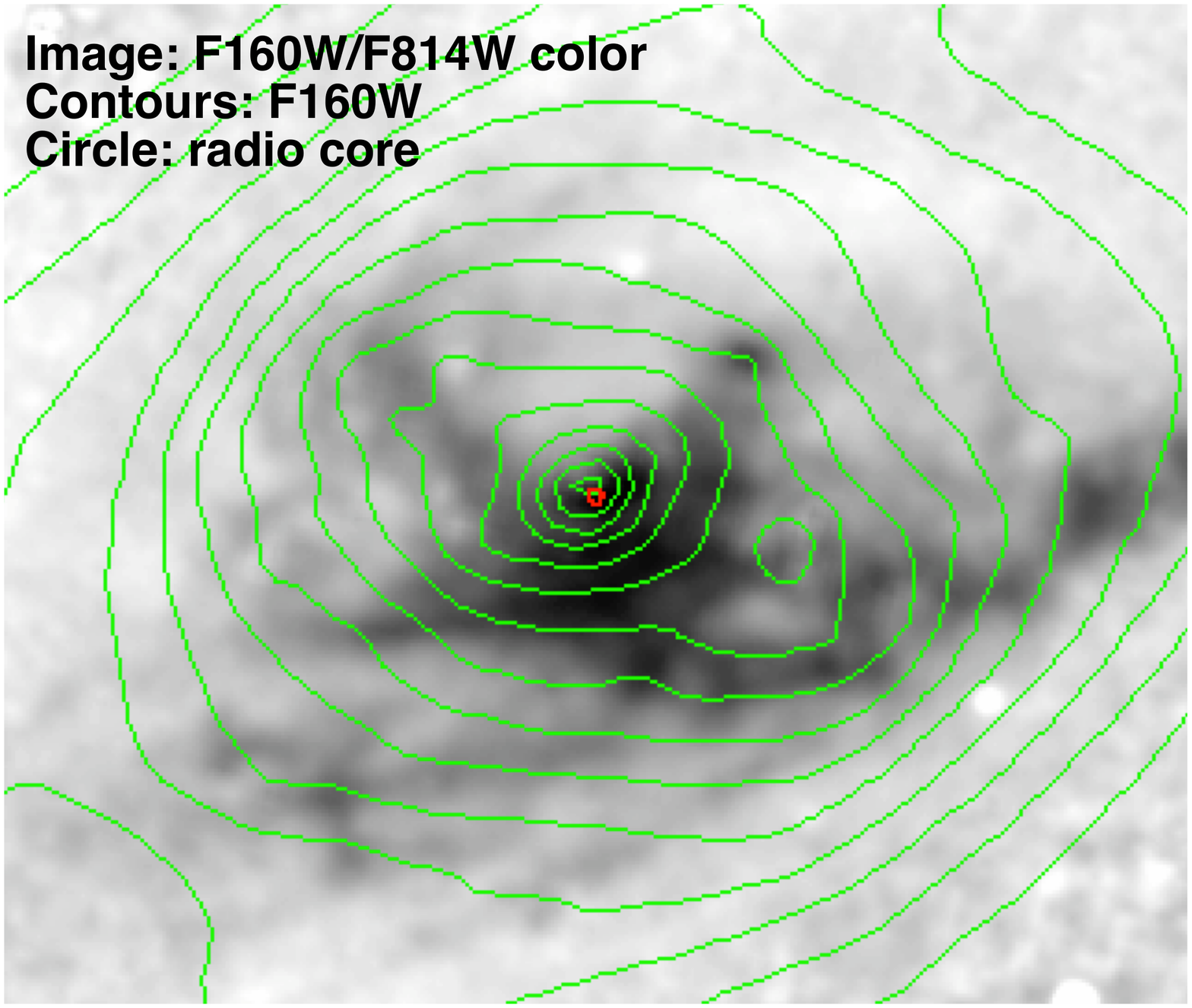}
\caption{Structures near the nucleus of IC 2497. The underlying image is the color map from Fig. \ref{fig-nuc}, now showing the innermost $12.2 \times 10.3$ arcseconds. Contours show 
the F160W flux, logarithmically spaced in intensity. The small circle shows the location of VLBI source C2, with a radius indicating the nominal accuracy of transfer between HST and radio reference frames.}
 \label{fig-nucoverlay}
\end{figure*}

Using intensity profiles of IC 2497 and a bright field star, we find that subtracting a nuclear point source 
brighter than $1.7 \times 10^{-18}$ erg cm$^{-2}$ s$^{-1}$ \AA$^{-1}$ at 
1.6 $\mu$m would leave a starlight profile with a central depression, so we take this as an upper limit to the emerging intensity directly from an AGN component.
This is a slightly less stringent limit than would be given by extrapolation of the X-ray power law detection by \cite{Schawinski2010}.
There is also a near-point source of emission in the F225W near-ultraviolet filter, but position registration shows its centroid to be displaced by 0.07 arcsecond
(1.6 WFC pixels) away from the dust lane compared to the peak in F814W, so its observed structure and flux are strongly influenced by the dust lane. A typical
reddening curve with $R=3.1$ would have an extinction at 2550 \AA\  of $\approx 8$ magnitudes for our estimated F814W extinction of 1 magnitude, so that the apparent UV nucleus
is either surrounding structure or is seen through patchy extinction that we have no good way to quantify.

The only way to hide a luminous AGN in this object without exceeding the observed FIR flux appears to be through beaming, as distinct from obscuration
by nearby material, as proposed for the emission-line filaments along the jet of 
Centaurus A by \cite{morganti92}\footnote{However, in Cen A, kinematic arguments suggest that shocks dominate the ionization in at least some parts of the jet structure.} The cone angle involved in IC 2497 would roughly span
55$^\circ$ in diameter to encompass all the detected [O III] emission, which is too broad to suppress the
putative nuclear emission in our direction to unobservable levels. Relativistic beaming would give this half-power width at a Lorentz factor
$\gamma \approx 2$, bulk velocity $v/c=0.86$, in which case the Doppler deboosting in our direction (about 125$^\circ$ from the axis) would be a modest
factor 0.3, leaving emission from the core very prominent. Thus, neither obscuration by a torus nor beaming offers an appealing explanation for why
Hanny's Voorwerp sees a luminous AGN which is virtually absent along our line of sight (providing further support for our interpretation in \citealt{Lintott09} and \citealt{Schawinski2010}).
 
 
\subsection{Spectrum and nuclear activity}

The STIS spectra of the nucleus of IC 2497 show line ratios characteristic of a Seyfert nucleus, on the low-ionization
side of the category in common diagnostic diagrams (Table \ref{tbl-nuclei}, Fig. \ref{fig-stisplot}). With reduced contamination from bulge starlight compared to our earlier ground-based data, the line equivalent widths are larger, shrinking the error bars due to uncertainty in the underlying starlight (particularly in H$\beta$ absorption; with the observed equivalent width of 25 \AA , corrections for underlying stellar absorption will not exceed 30\% (8 \AA ), and would lower the
derived [O III]/H$\beta$ and ionization level. The Doppler widths are substantial, enough to completely blend H$\alpha$ and the adjacent [N II] lines. We separate the contributions using Gaussian deblending with the line separations fixed. There may be a contribution to H$\alpha$ from a low-luminosity broad-line region, since the FWHM needed to fit the H$\alpha$+[N II] blend is twice that of the unblended [O III] $\lambda 5007$ line. Combining these features, the nuclear spectrum would
be classified as a Seyfert 2 or Seyfert 1.9 (depending on the detection of a BLR contribution at H$\alpha$).


\begin{figure*}
\includegraphics[scale=0.65,angle=0]{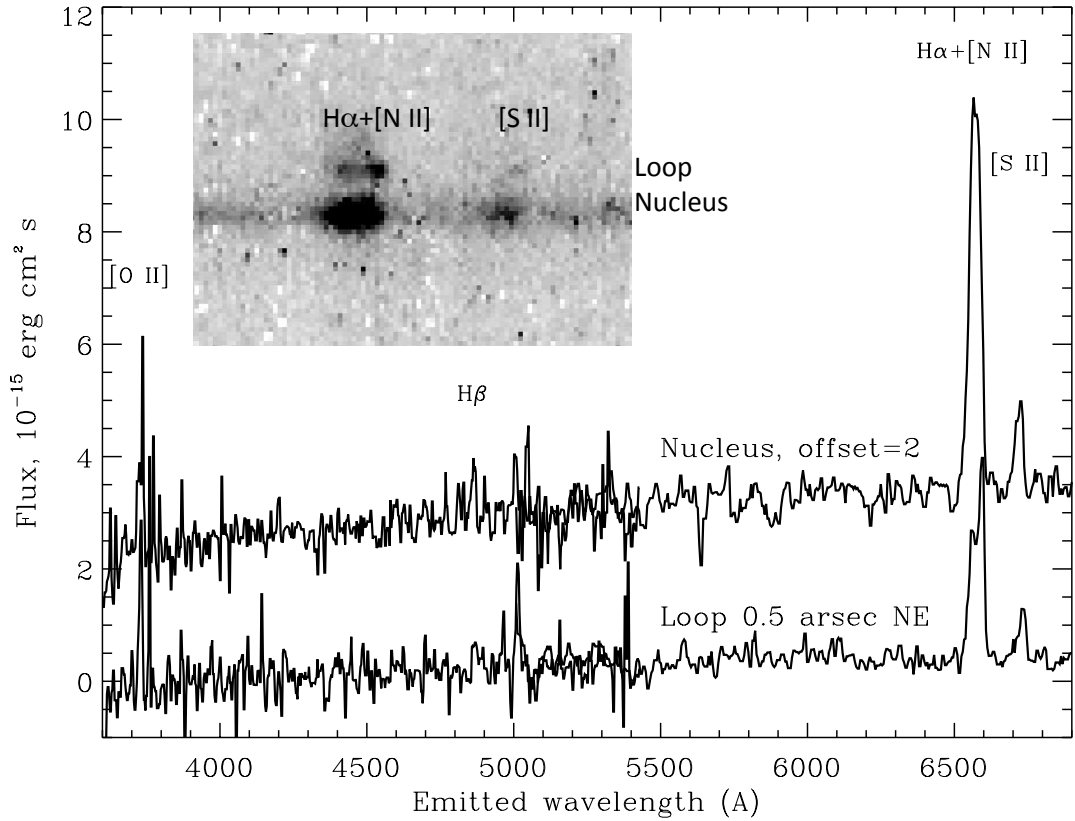}
\caption{Sections of the STIS spectra including the nucleus of IC 2497 and the emission region (``loop") 0.5" to the northeast. Each sums over a region $0.15 \times 0.20$ arcsecond. The nuclear spectrum is offset upward by 2 units for clarity.  For display purposes, the spectra have been filtered using a 3-pixel median to reduce the cosmetic effects of cosmic-ray events. This shows distinctions in 
emission-line ratios, and the linewidths in the H$\alpha$+[N II] blend, between the two
regions. The inset shows the two-dimensional red spectrum in the region of H$\alpha$, [N II] and [S II], again highlighting the
differences between the nucleus and adjacent emission-line loop}
 \label{fig-stisplot}
\end{figure*}

The two-dimensional profiles of [O III] and H$\beta$ show some ionization and
velocity structure close to the resolution of the spectra.

The line ratios let us derive an ionization parameter for photoionization, following the procedure in
\citet{Bennert}. We deredden the ratios taking the case B Balmer decrement of H$\alpha$/H$\beta$=2.87; if a distinct broad component of
H$\alpha$ contributes significantly, the dereddening correction would be smaller. For the nuclear emission region, the oxygen lines give log $U = -3.5$, 
somewhat lower than
our estimate of -3.2 in \cite{Lintott09}. The difference is due both to tighter bounds on H$\beta$, since
contamination from stellar absorption is much smaller, and better sensitivity and calibration accuracy
into the blue.

The nuclear emission region is small - FWHM  4 pixels along the slit (0.20"). Thus the
characteristic distance of the ionized region from the core is $ < 0.10$" (100 pc) if this
region is centered on the core source. This size measurement makes the constraints
on the ionizing luminosity of the nucleus from \citet{Lintott09} even stronger. In a straightforward scaling, the  gas near the nucleus is 150 times closer to the AGN than the Voorwerp, while the density
$n_e=560$ cm$^{-3}$ \citep{Lintott09} is 10--40 times higher. Thus the space density of ionizing photons is 600--2500
times smaller in the circumnuclear gas than it is upon reaching the Voorwerp. This remarkable mismatch is
nontrivial, since producing these line ratios from photoionization requires a radiation field which is as hard 
in spectral shape as typical AGN continua. An obscured AGN would not provide this; lines of sight sufficiently clear to see ionized gas with these line ratios would be sufficiently unreddened to reveal a significant fraction of the core luminosity in their ionization parameter

\subsection{Expanding circumnuclear gas\label{loopsec}}
A second distinct emission-line region appears in the STIS spectra, 0.5" to the northeast of the nucleus (485 pc in
projection at an angular-diameter distance of 198 Mpc). The H$\alpha$ ACS image shows this as a coherent loop or ring 
passing through the nucleus, of which only a partial arc is strong in [O III]; its definition can be improved by using the STIS
broadband acquisition image as a continuum estimate (Fig. \ref{fig-ring}).

\begin{figure*}
\includegraphics[scale=0.65,angle=0]{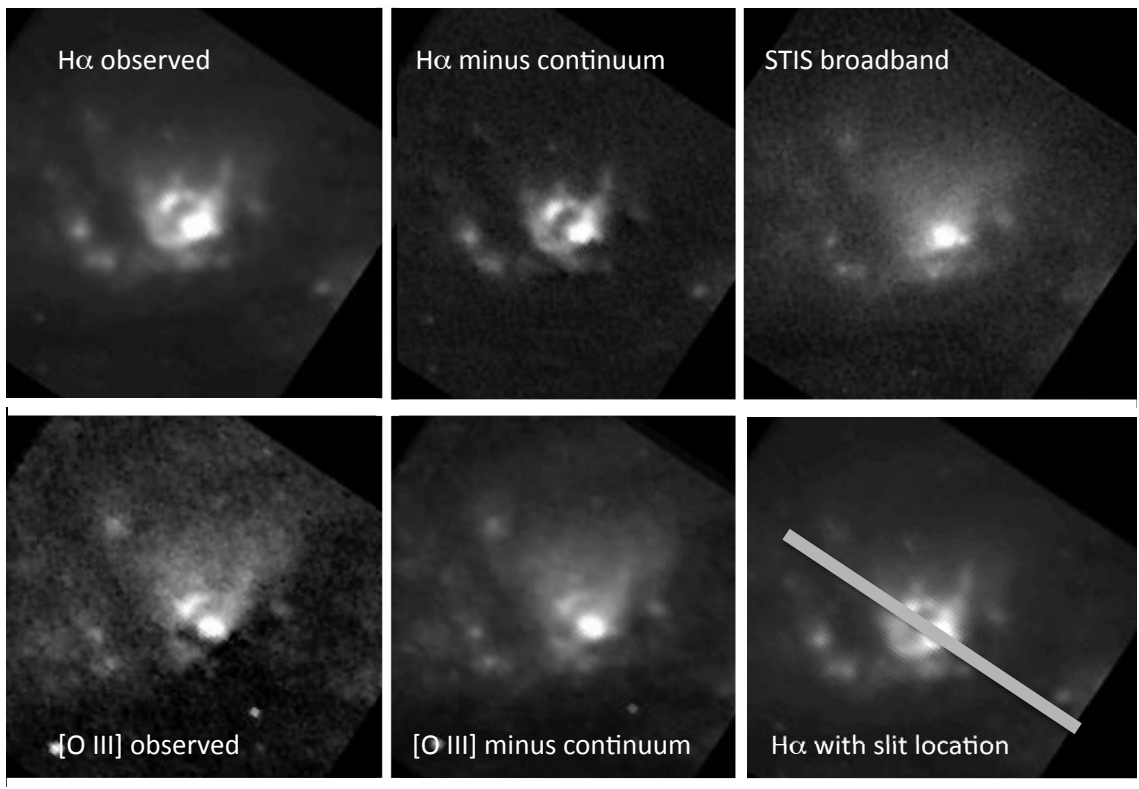}
\caption{Near-nuclear ring in H$\alpha$ and [O III] from the ACS ramp images, with the STIS broadband acquisition image used
for approximate continuum subtraction to show the emission structures more clearly. This was more successful at H$\alpha$ because of the better
wavelength match. The image shave been rotated for display to have north at the top. The bottom right panel repeats the ACS H$\alpha$ image,
superimposing the location and 0.2" width of the slit used for the STIS spectra.
The emission-line loop has a diameter $\approx 0.5$ arcseconds.}
\label{fig-ring}
\end{figure*}

Compared to the nucleus, this region has
larger redshift,  and somewhat higher levels of both ionization and excitation (Table \ref{tbl-nuclei}). The linewidth from the [N II]+H$\alpha$ blend is comparable to that from [O III] in the loop, in contrast to the nucleus, and the [O III] line width is smaller in the loop.

Can this structure be fairly described as a flat ring, or could it be a roughly spherical shell or bubble such as is seen on larger scales around
a few AGN? The surface-brightness profiles in both H$\alpha$ and [O III] are highly
symmetric across the loop, after subtracting a sloping background from inside and outside.The radial emission profile 
is not well resolved, with FWHM$\approx 0.15$" from the H$\alpha$ image (0.20" from the STIS spectrum after 
continuum subtraction) about the peak radius. Comparison with 
simple numerical models of a spherical shell shows emission inside the ring at a higher level than we observe; this
structure must be more like a flat ring than a complete shell. An additional emission-line
feature appears beyond the ring to the northeast, most obviously interpreted as a
component with strong [N II] at velocities gradually returning to nearly the central value
over a span of 8 pixels (0.32", 300 pc). This supports an interpretation of the ring as
a local kinematic structure within more quiescent material.

For an expanding ring viewed at an angle $i$ to its normal, the characteristic expansion age is given
from observables (line-of-sight expansion velocity $v_{los}$, apparent minor diameter $D_{app}$)
by  $T = D_{app} \cot i / v_{los}$.  For $v_{los} = 310$ km s$^{-1}$, this becomes  $1.5 \times 10^6 \cot i$ years. 
(Since our slit was not aligned with the loop as seen in direct images, we do not sample $v_{los}$ on the minor axis of the loop; this value could be larger
and the loop correspondingly younger). 
If the ring is in the plane of IC 2497, $i \approx 65^\circ$, so $T \approx 7 \times 10^5$ years. Expansion
perpendicular to the plane is ruled out by noting that it is projected on the far side of the plane
as marked by the dust lanes and redshifted relative to the nucleus. The closer a ring lies to the plane of the sky, 
the shorter its expansion age for a given radial-velocity span. Of course, more
complex nonradial motions and a nonlinear expansion history  are allowed by our very limited 
sampling of its velocity (making our age estimate likely an upper limit). 

The expansion timescale of this feature could be broadly comparable to the inferred time since the AGN faded; 
an intriguing possibility is that it was powered by a change in accretion mode to one in which
much of the energy emerged in kinetic rather than radiative form (``radio mode"). Such a
mode of accretion power has been discussed in connection with radio-loud AGN having very
weak continuum emission at higher energies (\citealt{Churazov}, \citealt{MH2007}).
In particular, \cite{MH2008} find that a kinetically-dominated mode dominates at low Eddington ratio, so that 
such a switch could indeed be associated with a drop in accretion rate.
 
The relatively large velocity dispersion of the emission lines in the ring suggests that the kinematics
are more complicated than simple expansion in the galaxy's plane. With $\sigma_v = 325$ km s$^{-1}$ from 
Gaussian fits, the internal spread is a large fraction of the systematic Doppler component,
yet the emission ring is narrower than would be expected from free expansion about this mean.
Interpretations might be that motions are largely constrained to be normal to the ring (as at an interface between 
an outflow and dense disk), or that the emissivity drops rapidly away from the ring so that high-velocity gas 
disappears from the spectrum before moving far away.

Applying the \cite{Bennert} fitting forms, the line ratios in this loop give log $U = -3.27$, somewhat
higher than the nucleus. With only a few strong emission lines detected, and lines profiles broad enough
to accommodate the shock velocities associated with its ionization level,
shock ionization remains a possibility for  this feature.

Lack of more detailed information on the physical conditions in the expanding loop limits our ability to estimate the energy
required to produce it. The blended [S II] doublet has a mean wavelength implying that the doublet ratio is near unity, taking the
line redshift to match [N II]  emission in the loop. At a typical temperature $10^4$ K, this gives electron density $n_e \approx 900$
cm$^{-3}$. Very crudely, if we take an astrophysically typical filling factor $10^{-4}$ and fully ionized gas distributed in a ring of thickness
comparable to its width, the mass involved would be 
$\approx 3000$ solar masses. The kinetic energy of such a mass in a ring expanding at
300 km s$^{-1}$ is then $\approx 3 \times 10^{51}$ergs. Compared to the ionizing luminosity needed
to ionize the distant gas in the Voorwerp, this could be supplied by only a small fraction of the energy output
when spread over even a few thousand of years. In fact, even compared to the estimated current luminosity of the
AGN near $10^{40}$ erg s$^{-1}$, this remains a tiny fraction of the total output over the lifetime of the ring.
These could be additional outflows at temperatures not sampled by our data, such as a hot interior analogous to the
Fermi bubbles in the Milky Way \citep{Fermi}; this situation would increase the energy requirements. beyond our estimate for the loop itself.

  
Possibly related to the origin of the ionized-gas loop is structure seen in the UV F225W image to the north of the nucleus. 
As shown in Fig. \ref{fig-uvcore}, there is a larger plume with some of the brightest emission at about the same position angle
as the H$\alpha$ loop. The brightest region near the nucleus occurs at roughly the same position angle as the middle of the loop.
The origin of this emission is poorly constrained; the filter is dominated by continuum processes, and
detection of H II regions elsewhere in IC 2497 confirms that the exposure is deep enough to see bright star-forming regions,
but there is no correlation between the UV and H$\alpha$ near the nucleus. This is not synchrotron emission from a radio
jet, because the observed jet goes in the opposite direction, so that any radio emission on this side would be far too weak be
produce detectable UV emission. Similarly, scattered light from an otherwise obscured and luminous AGN
would occupy such a small solid angle in this feature as to suggest that most of its radiation is absorbed near the nucleus, violating the
constraints on far-IR luminosity. The roughly radial UV structures are elongated roughly in PA $30^\circ$, which is not opposite either
the Voorwerp or the radio jet knots.
The data would be satisfied with a stellar population which is young, but still
old enough not to ionize any nearby gas. The dust lanes make it clear that the UV feature is projected against the far side of the
disk, and that any counterpart on the south side of the nucleus would be invisibly faint behind the dust.

\begin{figure*}
\includegraphics[scale=0.6,angle=0]{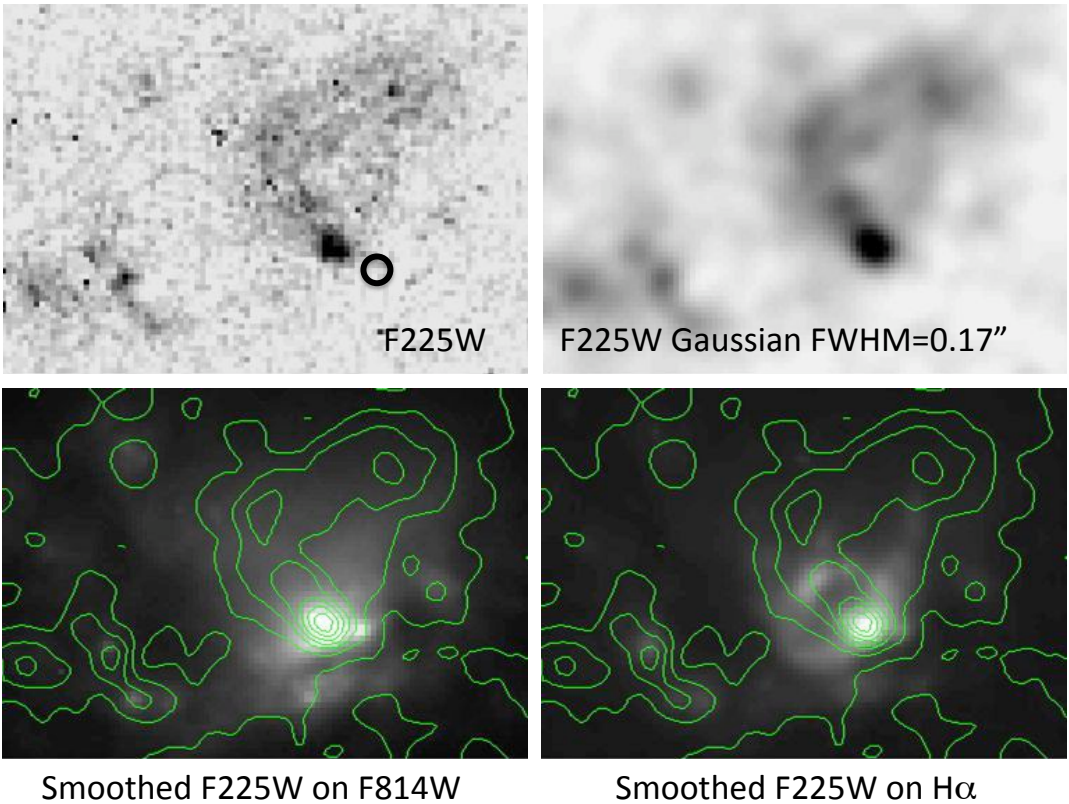}
\caption{Near-UV structure near the nucleus of IC 2497, in the WFC3 F2525W filter. The upper panels show the drizzle-combined
exposures ``as is" and with a gaussian smoothing of FWHM=0.17". In the lower panels, this smooth image is overlaid as contours on the
F814W and H$\alpha$ images to show the registration of various components. The circle overlaid on the ``as is" image shows the 
core VLBI component C2's position and approximate $2 \sigma$ position circle in the coordinate match. Each panel shows a region $66 \times 94$ pixels,
or $2.64 \times 3.8$ arcseconds. North is at the top.}
\label{fig-uvcore}
\end{figure*}

\subsection{Spectral energy distribution and luminosity}

Using data obtained since the preparation of \cite{Lintott09},
we can revisit bounds on the bolometric luminosity of the AGN in IC 2497.
We consider the X-ray results of Schawinski et al., and the 
recently released mid-IR values from WISE.

As noted by \cite{Lintott09}, the FIR parameter, estimating the 
total output from 42-122 $\mu$m from IRAS fluxes at 60 and 100 $\mu$m, gives a far-IR
luminosity of $6 \times 10^{44}$ erg s$^{-1}$ for IC 2497. With the addition of WISE detections \citep{WISE}
from near- to mid-IR, we can fill in the spectral energy distribution
with measurements tracing very hot dust (which turns out not to
be energetically important in this galaxy). Zero points of the WISE
magnitude scales were taken from \cite{Jarrett}. IC 2497 has
a near-IR spectral slope which does not indicate dominance by the AGN, using the
color criterion from \cite{Stern}; they present evidence that this would include even 
Compton-thick objects at high luminosity.

We have done a more detailed 
integration of the far-IR luminosity, combining IRAS and WISE measurements and
 interpolating between measured
bands with piecewise power-law shapes. Extrapolation beyond 100 $\mu$m
used a typical galaxy shape, a 20 K modified Planck law with emissivity 
scaling as $\lambda^{-2}$, and the whole composite spectrum was taken to peak 
at 150 $\mu$m. The resulting IR luminosity from 3.4-122 $\mu$m
is $8.0 \times 10^{44}$ erg
s${-1}$, about 25\% larger than given by the the FIR parameter. 
Extrapolation to longer wavelengths using our simple model would raise
the total to $1.0 \times 10^{45}$ erg s${-1}$. As usual, this is an upper limit
to reprocessed AGN radiation, since star formation also contributes
(particularly at longer wavelengths). The implied ionizing luminosity alone for the
AGN is twice this limit, and obscuration deep enough to hide the AGN so completely
in the optical would also include non-ionizing radiation (very roughly three times its
energy contribution for a typical radio-quiet AGN energy distribution. This means that
the most generous accounting for reradiation from absorbed AGN emission gives
a shortfall by a factor $\approx 6$. The magnitude of the shortfall is increased when
including an FIR contribution from star formation
in IC 2497, the lack of an AGN color signature in the mid-IR, and the 
much stronger limits from emission-line gas in its nucleus.

\section{Morphology of IC 2497}

IC 2497 is a very nearby quasar host galaxy, making its morphology of particular
interest. Combining the spiral structure of IC 2497 (which was already clear from SDSS images) and the evidence for a radio jet or outflow (\citealt{Josza}, \citealt{Rampadarath}), this is one of the very rare instances of an unquestionable spiral galaxy hosting large-scale radio jets (\citealt{Keel0313}, \citealt{Hota}).
Multiple circumstances may be necessary for such a rare combination -- a very massive
central black hole, triggering event for accretion, near-polar alignment within the host galaxy, and appropriate density of the local intergalactic medium to make the jets bright enough to detect..

The WFC3 F814W and F160W images show a bar, punctuated by bright H II regions.
This structure is stressed in the logarithmic mapping used in Fig. \ref{fig-galaxiesf814}, including IC 2497
and its spiral companion.
The prominent two-armed spiral structure is warped out of the plane of the inner disk; dust associated
with the foreground arm to the west and south appears in projection against a large area extending close to the nucleus. Associated dust lanes in the inner
disk and bar are projected close to the nucleus, motivating our examination of the position
of the nuclear radio source. 
Star formation is ongoing; there are luminous star clusters and H II regions in the disk and bar. In particular,
the H$\alpha$ structure seen to the southwest of the core by \citet{Lintott09} is resolved
into multiple H II regions with weak [O III], also seen in the Kitt Peak long-slit spectrum. Several disk clusters are both young enough and unobscured enough to be bright in the UV F225W image. Just to the northwest of the nucleus is a partial
loop in the UV, distinct from and larger than the emission-line ring feature (section \ref{loopsec}).

\begin{figure*}
\includegraphics[scale=0.6,angle=270]{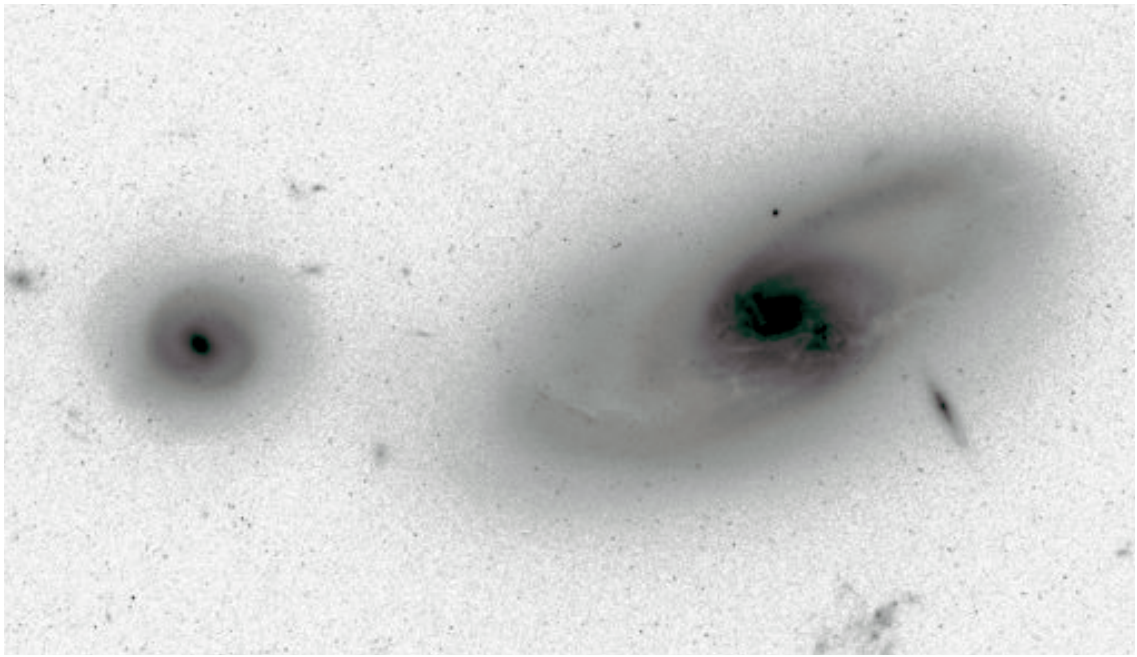}
\caption{I-band (F814W) WFC3 image of IC 2497 and its spiral companion to the east, displayed with a logarithmic
intensity scale. North is at the top; the region shown spans  $  64.3 \times  37.3 $ arcseconds.}
\label{fig-galaxiesf814}
\end{figure*}

Dust lanes associated with the near-side (warped) spiral arm cross the central region
nearly projected against the nucleus, as noted above. The companion galaxy just to the east 
lies at a closely matching redshift (section 2.5.2), making it a probable physical companion. However, its high degree of symmetry,
and modest luminosity (5.5 times smaller in the $r$ band) compared to IC 2497 itself, make it 
unlikely to have lost the  $\approx 9 \times 10^9$ M$_\odot$ of H I seen in the
curved tail to the south of both galaxies \citep{Josza}.

The symmetry of the companion, and radial-velocity difference near zero, place limits on the kind of encounter it might have undergone with
IC 2497.
Simulations varying  encounter parameters show that the least tidal damage for a given impact
parameter and mass ratio occurs when a galaxy undergoes a near-retrograde encounter, and
the greatest damage giving a warped disk will occur for an encounter inclined to the galaxy
plane but in its direction of rotation (e.g., \citealt{simatlas}). These can be broadly satisfied if the
relative orbit of the companion galaxy is inclined $30-45 ^\circ$ to the plane of IC 2497, and if
we see the companion as currently moving clockwise on the sky and toward us. Still, the large H I mass in 
the external features suggests that much or all of it came from IC 2497 or a now-disrupted companion, again at odds with the
usual situation in such unequal-mass encounters. Despite its proximity in position and redshift, this smaller spiral companion may 
thus not be directly relevant to the system's tidal history.

There are a few examples of similarly extensive tidal features with comparably large H I masses
\citep{Rogues}.
The Leo Ring has $\approx 2 \times 10^9$ solar masses of H I, and recent data
are compatible with a tidal origin (\citealt{Leo1989}, \citealt{Leo2010}). The much
more massive structure around NGC 5291 encompasses $6 \times 10^{10}$ M$_\odot$ in H I in
a $\approx 150$ kpc arc
around the interacting NGC 5291/Seashell Galaxy system (\citealt{Malphrus}, \citealt{Boquien}).
In each of these cases, there are interacting galaxies of comparable luminosity, and thus more
obvious donors for the extended H I, than in the case of IC 2497. One possibility is that IC 2497
itself is the aftermath of a major merger, one which had the appropriate geometry or mass
ratio to retain much of its gas. The clump of star-forming regions southwest of its nucleus
might, in this case, represent a remnant of the former companion (similar to the proposed
remnant of the companion responsible for the long tidal tail in the Tadpole system VV29 = UGC 10214;  
\citealt{Briggs}).

\section{Structure of Hanny's Voorwerp}

We now turn to the structure of Hanny's Voorwerp itself as revealed by the HST images. This includes the morphology of the
ionized gas and its ionization structure, and the presence of embedded young star clusters surrounded by H II regions.
An overview of the structure in emission lines may be seen in Fig. \ref{fig-color}, combining the [O III] and H$\alpha$ images in color.

\begin{figure*}
\includegraphics[scale=0.7,angle=90]{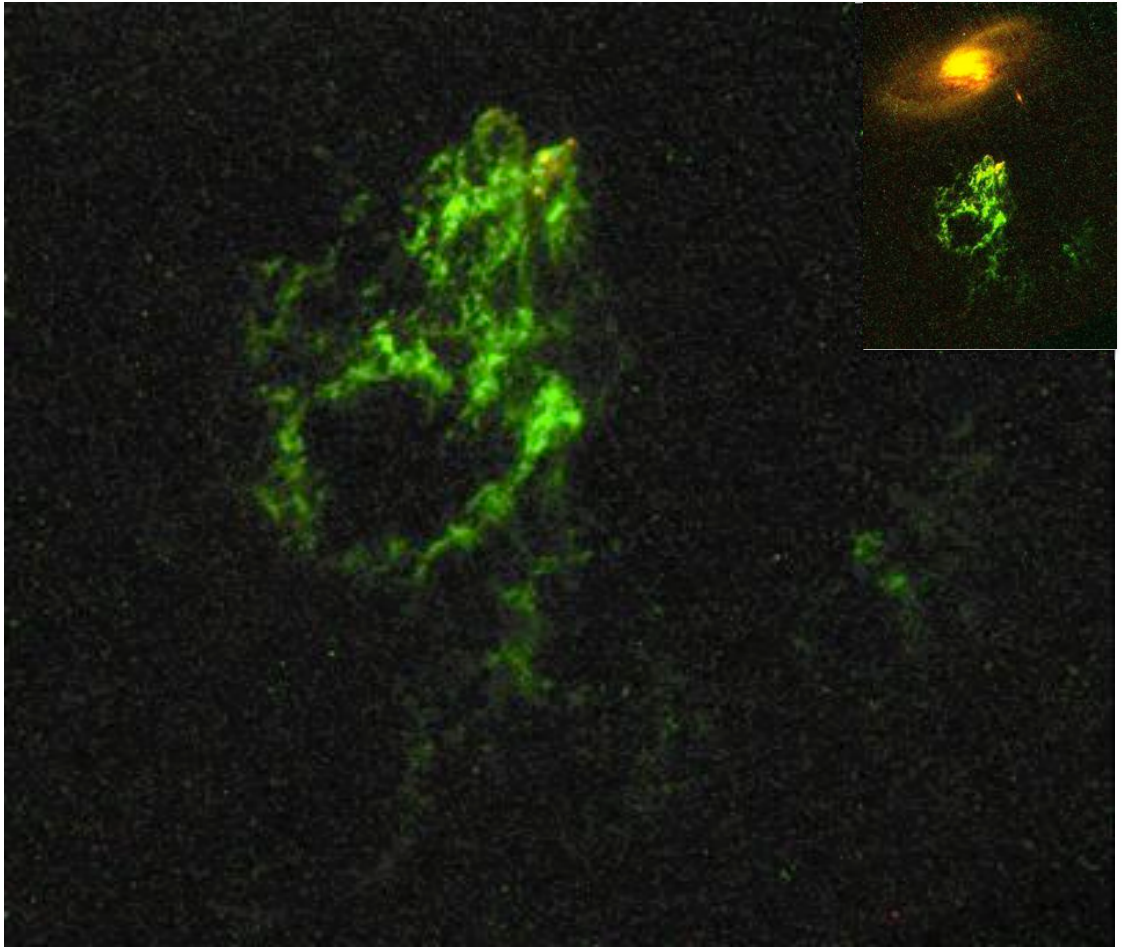}
\caption{Color composite of the emission-line structure in Hanny's Voorwerp. [O III] is mapped to green and H$\alpha$+[N II] to red,
with linear intensity scales set to equal contributions at [O III]/H$\alpha$=1. North is at the top and east to the left; the field of the
main image is $34.4 \times 40.4$". This includes all emission regions shown in ground-based images, and covers the full
monochromatic field width of the ACS ramp filters E-W. The inset is a similar display of a wider field to show the relative location of IC 2497. }
\label{fig-color}
\end{figure*}

\subsection{Emission-line structure}

The gaseous structure of Hanny's Voorwerp is best traced by the strong [O III] emission line. As shown in Fig. \ref{fig-o3}, it is strongly filamentary, with structures on all scales down to the ACS resolution limit. Among major features are the prominent ``hole" surrounded by bright filaments, a small-scale loop or shell at the
northern end toward IC 2497, a region with embedded star clusters to the
northwest, and outlying regions of low surface brightness spanning much of the east-west extent of the filter's monochromatic field.
At the low densities derived from the [S II] line ratio, the recombination timescale $(\alpha n_e)^{-1}$ (where $\alpha$ is the
recombination coefficient) is long compared to the light-travel time across all 
but the largest of these structures. For hydrogen, $n_e <  50$ cm$^{-3}$ \citep{Lintott09} gives $t_{rec} > 2400$ years under nebular ``Case B" \citep{AGNAGN}.
Since the recombination coefficient $\alpha$ differs for various species, when the ionizing radiation is turned off, we would see different ionic species recombine
and disappear from the spectrum at different times (as calculated by \citealt{Binette1987}). For instance, O$^{2+}$ recombines $\approx 100$ times faster then H$^+$
\citep{Crenshaw2010}. The density dependence of these values means that we might see preferential fading of dense regions after a decline in ionizing
flux, which would act to flatten the intensity profile observed from a centrally concentrated cloud. This effect would change prominent line ratios such as [O III]/H$\alpha$
in the same sense as  changes in ionization parameter due solely to density, changes which we do not observe (section \ref{sec-ionpar}); both processes are likely overshadowed
by the role of unresolved fine structure in the emission-line gas.
 
 \begin{figure*}
\includegraphics[scale=0.5,angle=0]{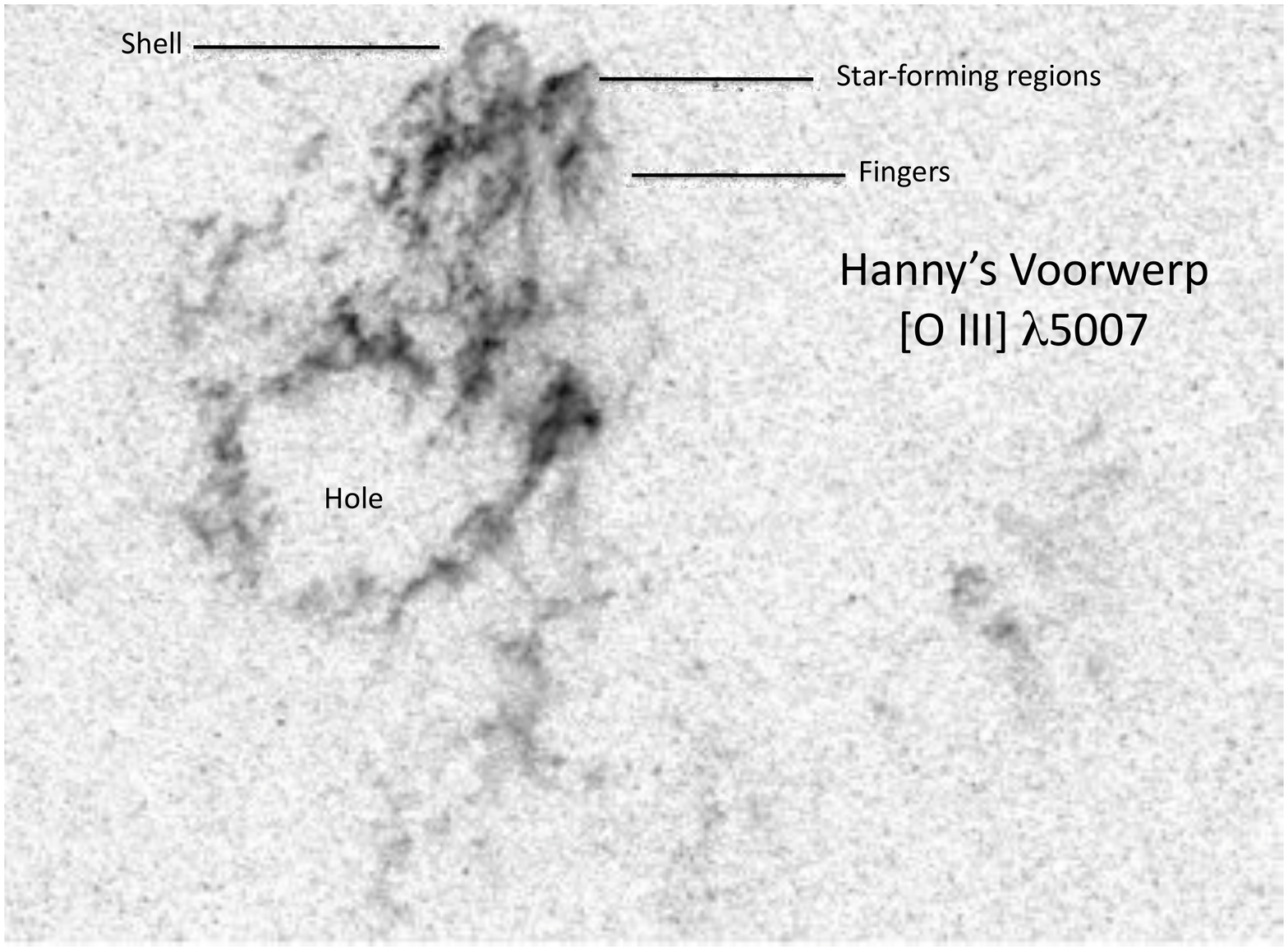}
\caption{[O III] structure of Hanny's Voorwerp from the ACS ramp filter, showing faint outlying emission regions. The intensity scale uses a logarithmic
intensity mapping starting slightly below the sky level, giving sensitivity to structure at a wide intensity range much like that recommended by
Lupton et al. (2004). The region shown spans  $26.6 \times 36.3$ arcseconds, with north at the top. Some features discussed in the text are indicated.}
 \label{fig-o3}
\end{figure*}

 Earlier imaging, in particular the WIYN $V$-band data,  led us to speculate on features possibly resembling bow shocks. This resemblance is less
 striking in the HST data, but one structure of particular interest may show interaction with an external medium at the northern end of Hanny's Voorwerp.
 This is a nearly circular shell or loop, 1.9" in diameter and 0.6" thick (Fig. \ref{fig-o3}). However, it could also result from a local explosive event, or be
 a chance configuration of filaments. We do not yet have kinematic data to test any of these possibilities. This ring does have areas of relatively 
 low [O III]/H$\alpha$ around much of its circumference, lower than found elsewhere except for the star-forning regions to its west (section 5.3).
 However, we do not see continuum sources here such as mark young star clusters in the western region.
 	
If the pattern of ionized gas traces an ionization cone, it is a ragged one. There are filaments of low surface brightness detected on either side
of the obvious structures, and the bright area would trace a rough cone only if the gas is oriented in three dimensions with
significant depth along the line of sight.
					
In the northern part of the Voorwerp, there are filaments which are edge-brightened on the side toward IC 2497, a situation which does not
persist farther to the south. These filaments include a region where we see recent star formation (section 5.3). The combination of morphology
and association with young stars, whose formation could be triggered by compression, fits with a picture in which these structures in the
gas are interacting with an outflow from IC 2497, such as is shown in this direction by the Westerbork continuum data from \cite{Josza}.
In this respect, an interaction has occurred analogous to that seen in such systems as
Minkowski's Object, and star-forming regions adjacent to the inner jet of Centaurus A. In contrast,
what appears to be a robust jet/galaxy interaction in 3C 321 has not
led to a starburst response \citep{Evans}, so the properties of both outflow and target gas
play roles in the outcome. 

The low density measured from [S II ] lines, and the modest role of the outflow in shaping 
the morphology of Hanny's Voorwerp, indicate that this wind from IC 2497 has a low ram pressure. 
In contrast, both the star-forming regions and H$\alpha$ emission in Minkowski's Object show
structures indicating strong interaction with the radio jet, with filaments in the downstream direction \citep{Croft2006}.
We see this only to a very limited extent, and only in confined regions, in Hanny's Voorwerp,
as shown in the [O III] image (Fig. \ref{fig-o3}). 
The best morphological case for such interaction - gas being entrained by an outflow - is in the series of
``fingers" south of the star-forming regions. They do not point to a single point, but do point roughly away from the galaxy,
arise from a single emission-line feature, and align with the star-forming area and the nucleus of IC 2497,
so that entrainment makes
sense if the outflow has enough ram pressure to overcome their internal pressure. This would suggest that the
outflow has a typical ram pressure comparable to the internal pressure in the emission-line gas, so that local variations
make the difference between significant and negligible roles for entrainment.

The star-forming regions and the emission-line ``fingers" are oriented in position angle $198^\circ$ from the nucleus
of IC 2497 and span a projected cone angle of width $\approx 10^\circ$, compared to the PA $209^\circ$ of the inner jet as traced by the VLBI core and bright knot \citep{Rampadarath}. It is thus plausible that these traces of interaction with an outflow show us a narrow stream which is radio-bright only in its inner
kiloparsec. This narrow outflow would be substantially misaligned with the emission-line hole (section 5.1.1), making its origin
in the same outflow unlikely

\subsubsection{The emission-line hole}
A prominent feature of the Voorwerp, noted by \cite{Lintott09} and indeed visible in the SDSS $g$ image,
is the dark ''hole" in its southeastern part. Potential explanations include passage of a narrow jet, an explosion, a shadow cast by an intervening cloud of very large column density or happenstance in distribution of ionized filaments. The new ACS
data address these only in a negative sense - revealing no structural traces of why this is here. 
The ionization level does not increase toward its edges (section 5.2.2), but some filaments do seem to go around its edges rather than crossing its edge and vanishing.
An explosive origin seems unlikely, since our data show no X-ray emission from any any hot bubble that would remain,
excess ionization at its edges, or (within our limited spectroscopic information) a kinematic signature of much larger radial velocities at its edges.

If the hole traces the location of a (past) radio jet, it has left no trace in the emission-line material. Filaments do not turn ``downstream" within the hole, as is seen in, for example, Minkowski's Object, where jet interaction is clear. The best estimate of the direction of a radio jet comes from the
MERLIN and EVLBI observations of the core and inner knot by \cite{Rampadarath}, which
are aligned along position angle $209^\circ$. In contrast, the direction from the nuclear component to the center of the hole lies along PA $181^\circ$, with the entire structure ranging from $173-190^\circ$. 
A jet would have to curve by at least $28^\circ$ in projection to match the two locations.

A somewhat similar ``hole" structure, within a larger region of bright filaments, is seen among the emission-line patches along the northeastern jet of Centaurus A. In the images by \cite{GrahamPrice} and \cite{KeelCenA}, the ``necklace" region likewise
shows a roughly circular region surrounded by emission filaments. For comparison, this region is illustrated in Fig. \ref{fig-cenaloop}. In contrast to Hanny's Voorwerp,
however, the ionization of these filaments in Centaurus A is attributed (mostly) to shocks, whose kinematic signature
appears in the wide velocity range and broad lines of individual features \citep{CenAshocks}. In
Cen A as well, it is not clear whether the ``hole" has direct physical significance, or results only from the arrangement of denser filamentary regions around it.

An interpretation as a shadow from a cloud near the AGN does not violate any of the data, but we cannot make a definite statement on the origin of this
feature.

\begin{figure*}
\includegraphics[scale=0.7,angle=0]{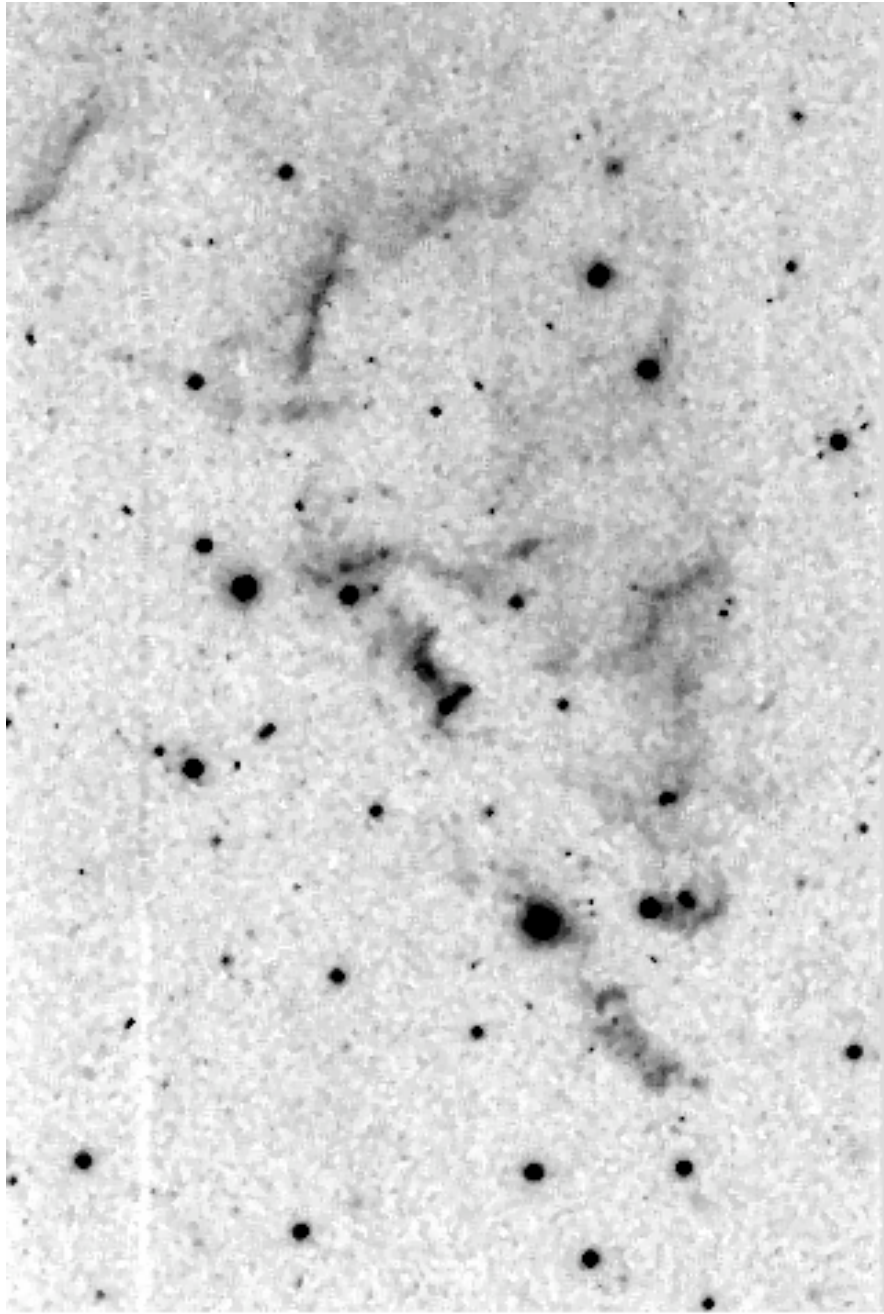}
\caption{Region of the Centaurus A emission-line jet morphologically resembling the ``hole" in Hanny's Voorwerp. This image from the ESO/MPI
2.2m telescope used an H$\alpha$ filter with FWHM= 36 \AA , from \cite{KeelCenA}. 
This ``necklace" region
has its brightest emission located near (2000) $\alpha$=13:2:28.5, $\delta$=-42:50:02. The image spans $95 \times 143$ arcseconds, with north at the top.
At a distance of 3.8 Mpc \citep{Harris2010}, the field corresponds to $1.75 \times 2.6$ kpc,  an order of magnitude smaller than the structure in
Hanny's Voorwerp. }
\label{fig-cenaloop}
\end{figure*}

\subsection{Ionization Structure}
	
The small-scale structure revealed by the ACS images allows us to probe the ionization structure within the gas, and improve our
limits on the AGN luminosity required to ionize it. This situation, with the same external radiation field at essentially the same
intensity impinging on large areas of gas, allows us to apply basic photoionization and recombination principles to probe
both the ionizing source and small-scale structure of the gas. 

\subsubsection{Ionizing luminosity}

As noted by \cite{Lintott09} and \cite{Keel11}, the surface brightness in recombination lines provides a lower limit to the luminosity
needed to maintain the ionization. The new images reveal bright features that were smeared out in our ground-based images, 
increasing the peak surface brightness and derived ionizing luminosity at each projected radius from the galaxy core. We have
considered several ways to evaluate this peak surface brightness. It is straightforward to find the peak pixel values, but
these could be affected by overlap of multiple structures or seeing a feature which happens to be elongated along the line
of sight.  
To allow for the effects of overlapping structures, or
dense regions which happen to be elongated along the line of sight and therefore appear brighter than a simple calculation
for a sheet or spherical structure would indicate, we consider two measures of the brightest regions in
histograms of H$\alpha$ surface brightness in several bins of
distance from the center of IC 2497 (Fig. \ref{fig-hahistograms}). 
The values labelled Max in Fig. \ref{fig-hahistograms} are for the highest contiguously populated bin in H$\alpha$
surface brightness, while the values given as 99\% are the 99th percentile in surface brightness among pixels more than
$2 \sigma$ above the sky level. (Other measures, such as 1\% of peak value, show similar behavior). As expected, the
lower limits to ionizing luminosity are higher when we see finer structure; the average derived from the four plotted
zones is $L_{ion}  > 7 \times 10^{45}$ erg s$^{-1}$. In addition, while the peak surface brightness by
each measure declines for zones projected farther from IC 2497, this decline is shallower than the $r^{-2}$ expected for
identical gas parcels illuminated by the same source. This mismatch could be explained either through geometry or through
history of the ionizing source (systematic changes in characteristic gas density are disfavored by the ionization structure). 
Specifically, the quantity $r_{proj}^2 f$ for each flux measure $f$ increases monotonically with
$r_{proj}$ (with one slight violation between the outer two zones in the 99\% value). A geometric explanation would have the
gas lying close to a plane which is highly inclined to the plane of the sky and not radial to IC 2497, so that the mapping between $r_{proj}$ and $r$ 
differs for each zone. For example, if the southernmost zone had $r_{proj} = r$, the gas would lie in a plane tilted by at least $45^\circ$
to the plane of the sky.

\begin{figure*}
\includegraphics[scale=0.6,angle=0]{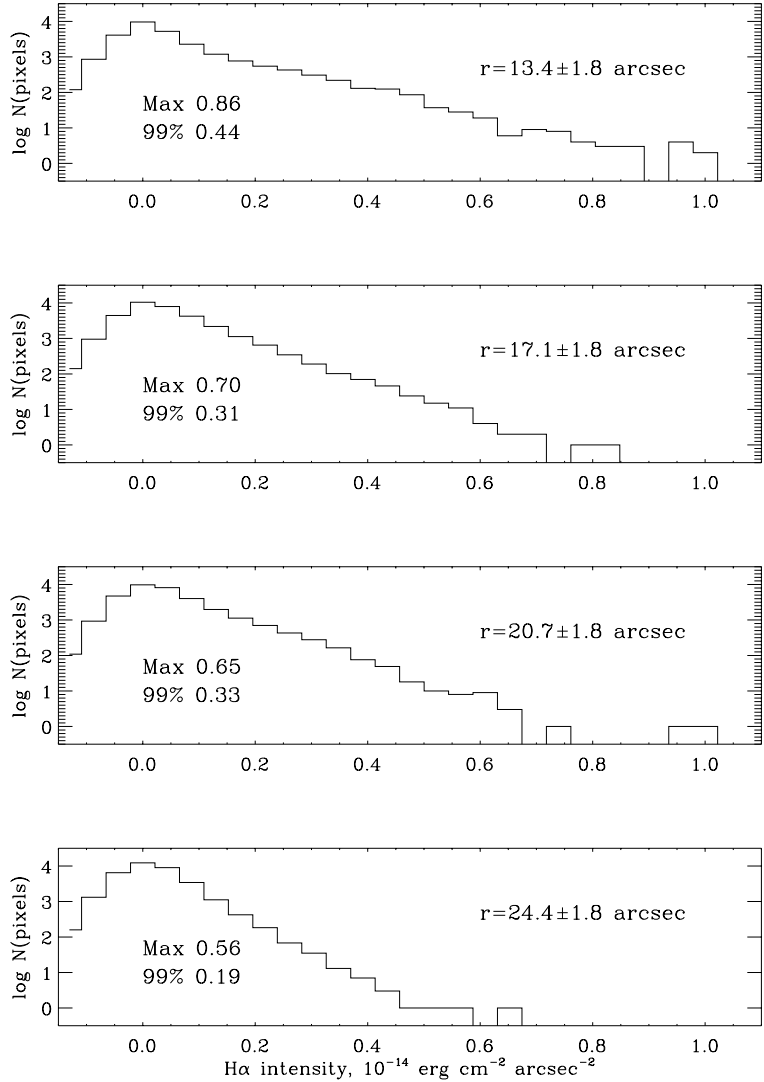}
\caption{Histograms of H$\alpha$ intensity for pixels in regions at various projected distances from the center of IC 2497. The area around the
star-forming regions with strong H$\alpha$ from local ionizing sources is omitted in this comparson. We estimate the
peak H$\alpha$ surface brightness, and by implication the minimum required ionizing luminosity, from features of each distribution. The panels list
two such values - the highest  contiguously populated bin, and the intensity at the 99th percentile among pixels more than $2 \sigma$ above 
the sky level. For each of these, the decline with projected radius is slower than $r^{-2}$. If the gas density distribution is constant across the
cloud, this would be seen if the gas distribution is strongly
tilted to the plane of the sky, or the ionizing source dropped in luminosity by a factor $\approx 2$ over the delay times sampled in this region.}
\label{fig-hahistograms}
\end{figure*}

Alternatively, the behavior of H$\alpha$ surface brightness with projected radius could result from long-term variations in ionizing 
luminosity, reflected through different values of the light-travel time delay. For a gas distribution viewed ``face-on" (in the plane
of the sky), the ionizing luminosity would have decreased by a factor $\approx 2$ over a timespan of  40,000 years. The uncertainty in 
interpretation here points up the importance of the system geometry; this might be addressed by kinematic mapping and modeling
of the entire H I stream.

\subsubsection{Density and ionization structure\label{sec-ionpar}}

Point by point in a nebula, the surface brightness in a recombination line scales with the emission measure
$$ EM = \int {n_e}^2 dl .$$
We have an independent tracer of changes in $n_e$ through the ionization parameter $U$; in this case,
small-scale variations in $U$ must trace changes in local density since adjacent pixels are at essentially the
same distance from the ionizing source. If the features seen are due only to density contrast, surface brightness
should correlate negatively with $U$, as deduced from emission-line ratios. We see at most a very mild
anticorrelation, so weak that most of the structure we see must result from changes in line-of-sight depth (or equivalently,
number of comparable features projected along the line of sight). To put this on a concrete basis, we use the
slope of the relation between $U$ and [O III]/H$\beta$ obtained for an AGN spectrum by \cite{Netzer90}, scaled
to [O III]/H$\alpha$ with an intrinsic Balmer decrement of 2.9. 

The [O III]/H$\alpha$ line-ratio
map (Fig. \ref{fig-o3ha}) shows an interesting variety of structures. There are small, discrete regions
of low [O III]/H$\alpha$, associated with the continuum objects in the northern part of the Voorwerp.
We identify these as star-forming regions, possibly triggered by compression of the gas
(Section \ref{starform}). Beyond this, the excitation is not well correlated with emission-line structures in the object (Fig. \ref{fig-uvsha}); the regions plotted
are identified in Fig. \ref{fig-o3label}. Ionization level is
not correlated with H$\alpha$ surface brightness (which is driven by density),
[O III] surface brightness (which combines ionization parameter and density), location within a filament center-to-edge, or location near the ``hole".  The highest line ratio
values occur
in several broad, roughly parallel stripes crossing the object, at a skew angle to the projected direction of the IC 2497 nucleus.

\begin{figure*}
\includegraphics[scale=0.6,angle=0]{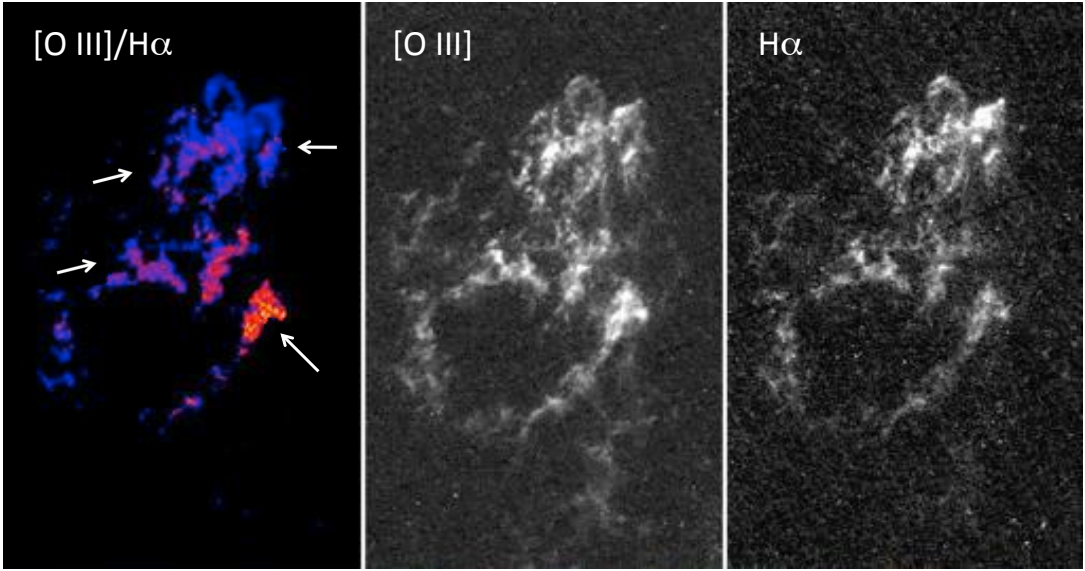}
\caption{Line ratio [O III] $\lambda 5007$/H$\alpha$ from the ACS ramp filters, masked
at low surface brightness where cosmic-ray artifacts dominate. The [O III] (center) and H$\alpha$ (right) images are shown for comparison. 
Several discrete regions of weak [O III] coincide with continuum objects, possibly star-forming regions. Outside these regions
the excitation level is not well correlated with surface brightness,  location within a filament, or distance from the ``hole". The highest excitation occurs in several broad, skewed strips crossing
the Voorwerp; the white arrows on the left panel mark rough locations of the clearest such strips. The color scale runs from zero to 8.0 in the line ratio.}
 \label{fig-o3ha}
\end{figure*}

\begin{figure*}
\includegraphics[scale=0.92,angle=0]{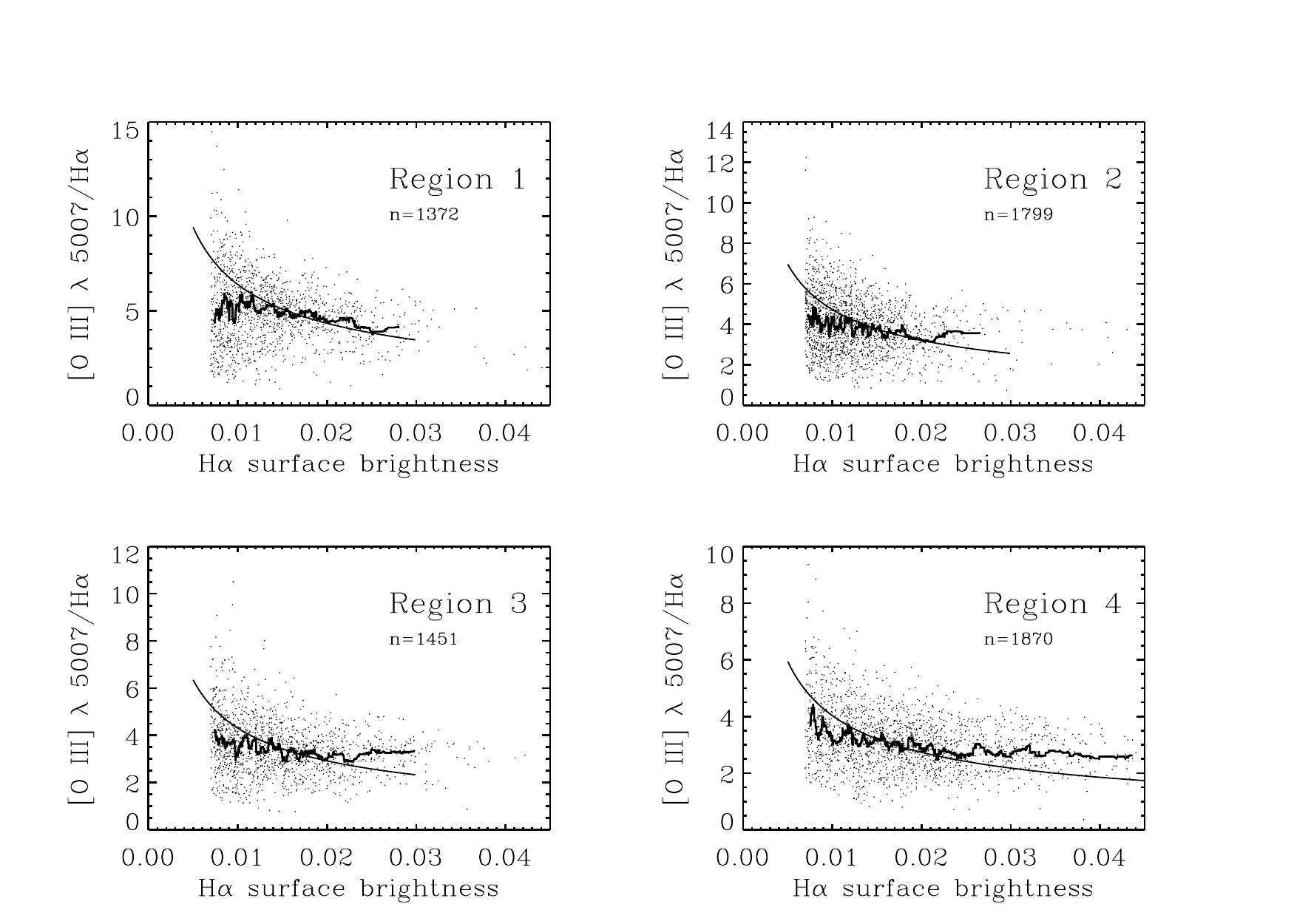}
\caption{Surface brightness-ionization behavior for four regions in Hanny's Voorwerp.
Individual pixels in the drizzled images are shown, with a $3 \sigma$ cutoff in H$\alpha$ (as used in the [O III]/H$\alpha$ mapping). The jagged
line is the running median across 51 points, shown where at least 37 points
from each end of the distributions. The smooth curves show the expected
behavior if the observed surface-brightness changes are due entirely to
density effects; the H$\alpha$ surface brightness would sample ${n_e}^2$.
In contrast, the [O III]/H$\alpha$ ratio changes with ionization parameter $U$ and therefore with density, since all pixels see the same ionizing spectrum (and nearly the same intensity within each region). Masking
for H$\alpha$ S/N$>3$ produces a slight upward bias in the distribution at small values, most apparent in the lower left corner for region 4. Photon
statistics contribute significantly to
the scatter of points at a given H$\alpha$ intensity, especially at low values. Each panel is labelled with the number of pixels shown.
H$\alpha$ surface brightness is given for simplicity in ADU second$^{-1}$.
One such unit corresponds to $1.45 \times 10^{-13}$ erg cm$^{-2}$
s$^{-1}$ arcsec$^{-2}$.}
\label{fig-uvsha}
\end{figure*}

\begin{figure*}
\includegraphics[scale=0.6,angle=0]{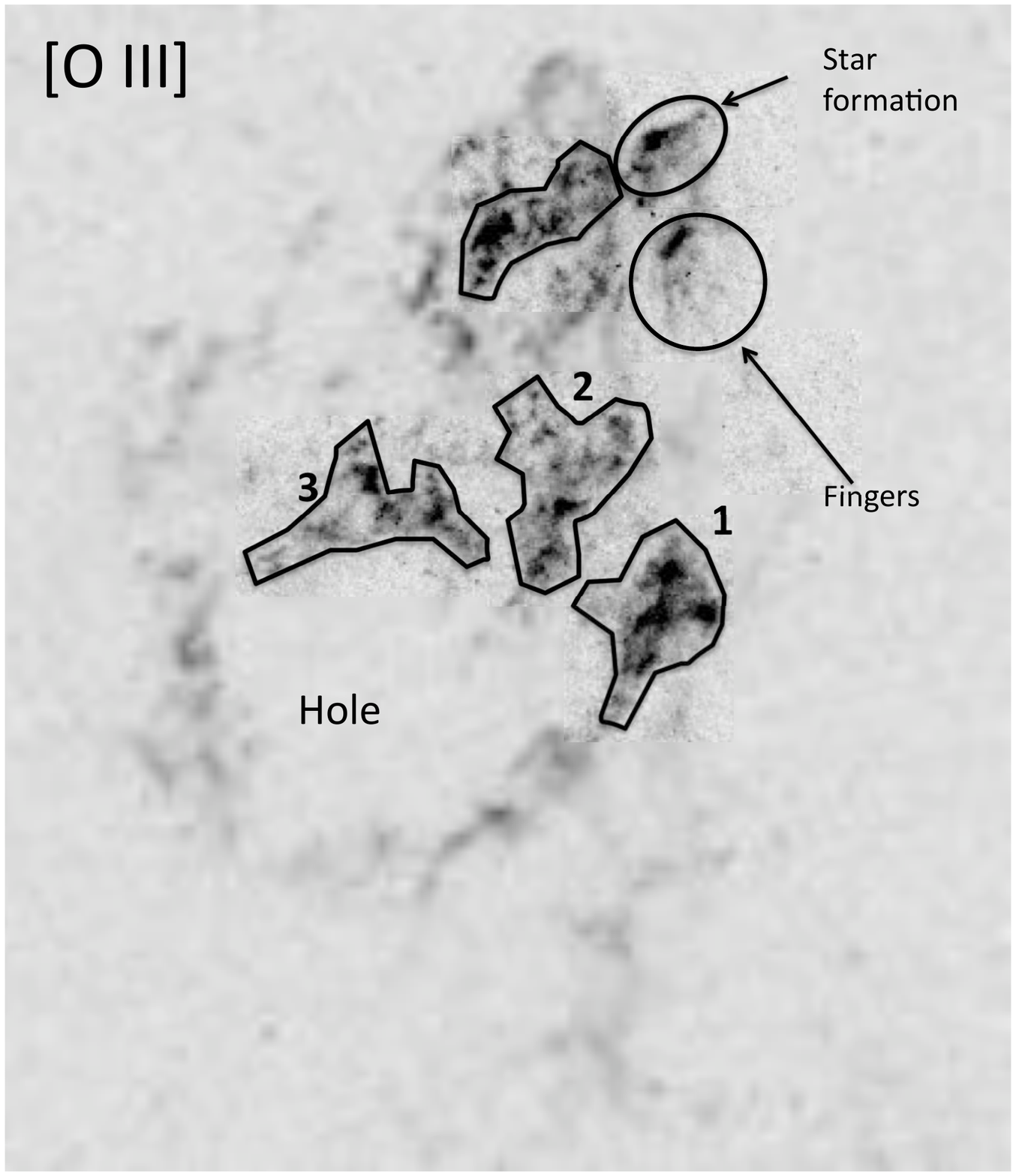}
\caption{[O III] image of the brightest parts of Hanny's Voorwerp, showing the four areas for which the ionization-surface brightness relation is
examined in Fig. \ref{fig-uvsha}. Other areas of interest are also marked for reference.}
\label{fig-o3label}
\end{figure*}

In the idealized case of a matter-bounded nebula, the surface brightness SB in a recombination line (such as the Balmer series) from a given volume
follows SB $\propto {n_e}^2$, while the ionization parameter $U \propto {n_e}^{-1}$.  If we are seeing a collection of
isolated and internally uniform clouds or filaments, we would expect to see $U \propto SB^{-1/2}$. Pixel by pixel, we see a much weaker, marginal average relation.
At least one of these simplifying assumptions does not apply. The relative strength of [O I] suggests that there are denser regions within the gas,
since this transition is favored in partially-ionized zones where O$^0$ and H$^+$ coexist. One way to account for the decoupling of surface brightness and
$U$ would be for the gas to have spatially unresolved fine structure (on scales smaller than the ACS resolution limit of $\approx 100 $ pc), so that the surface brightness we observe represents more closely the number of
clouds and filaments in a particular direction than their mean particle density. These might be analogous to the magnetically-confined structures inferred
by \cite{Fabian2008}; since we find that the outflow from IC 2497 has played only a minor role in the structure of the gas we see, an implication would be that the
rest of the H I tail is similarly structured, independent of action of the AGN. This fine structure means that we cannot detect any differential
recombination as the ionizing source fades; if a region with a density gradient were spatially resolved, we would expect to see species with faster
recombination timescales, such as O$^{+}$, fade faster than longer-lived ones such as H$^+$ \citep{Binette1987}.
Since the detailed
structure, in particular lack of filaments radial to IC 2497 in most of the object, indicates that reshaping of the gas by the galaxy's radio outflow has been modest, this suggests that the rest of the H I tail in this system is similarly filamentary. Gas in tidal streams may retain fine structure far beyond the spatial resolution of
current data.

These stripes of higher excitation (and almost certainly higher ionization) have projected width 0.7-1.2" (700-1200 pc). In the ionization-echo picture, these could represent periods of
higher ionizing flux, and would be significantly smeared by the recombination timescale $t_{rec} \approx (\alpha n_e )^{-1}$ for recombination coefficient $\alpha$.
As noted above, the emission-line limit on density $n_e < 50$  cm$^{-2}$ sets this timescale to
be 2400 years or more, scaling with $n_e ^{-1}$ \citep{Lintott09}. This light-travel time 
$c t_{rec}$ translates to 0.7"
or more spatially, so that the fine structure we see must be dominated by structure in the gas rather
than ionization history, and the filaments must be physically about as thick as they appear. This fits with the H I column density
if the ionized region is matter-bounded, and not optically thick in the Lyman continuum; thus
the lower bound on ionizing luminosity from recombination and energy budget from \cite{Lintott09} should indeed be higher.

If these stripes do represent ionizing outbursts, they suggest a geometry for the gas - in the light-echo
geometry, the intersection of the light-time ellipsoid with the ionized cloud is skew from the
direction to the nucleus when the surface is tilted both to our line of sight and to the radius
vector toward the ionizing source. 
 
We have little direct information on the three-dimensional geometry of the Voorwerp. The small
number of distinct filaments may indicate that it is geometrically thin, rather than being only
a thin ``skin" of ionized gas on the inner edge of a much thicker H I structure. Since the dominant
process in its excitation is photoionization rather than shocks \citep{Lintott09}, we expect there to have been little mass motion associated with its production, so that if it is thin, so is the H I stream. The
mean H I column density in this region is $ 8 \pm 2 \times 10^{19}$ atoms cm$^{-2}$ \citep{Josza}. The depth into which an external UV source would ionize this cloud depends on both the local density
and duration of exposure, which are poorly constrained. Emission-line diagnostics \citep{Lintott09} 
give an upper limit to the density from the [S II] line ratio, and energy balance shows that at least 
1/4 of the impinging ionizing photons are absorbed.

Where does the Voorwerp lie in distance, with respect to IC 2497? Escaping radiation would be more likely
near the poles of IC 2497, in view of the dusty disk features seen in our images,
and a minimum distance would give the most conservative ionizing luminosity. These factors suggest a somewhat longer time delay than the projected separation. If the ionized region is polar to IC 2497, its separation from the nucleus would make an angle $\theta \approx 125^\circ$ to the line of sight (about  
$35^\circ$ ``behind" the plane of the sky). The true distance between the nucleus of IC 2497 and the Voorwerp would then be $r_{proj} /\sin \theta$. The difference in light-travel time would then be 
$$ \Delta t = {{r_{proj}} \over {c \sin \theta}} (1 - \cos \theta ) \eqno(1) $$  
\citep{Keel11}. For
$\theta = 125^\circ$, this is greater than the plane-of-the-sky component of time delay $r_{proj}/c$ by a factor 1.9, spanning 97,000--230,000
years from the inner to outer regions of [O III] emission. It is possible that the 
entire emission-line structure forms a fairly thin, wrinkled sheet roughly perpendicular to the
incoming photons; in this case, the stripes of highest ionization level might record the same
peak in core luminosity. However, this geometry would require systematic changes in characteristic
density of the (unresolved) emitting structures to fit with the changes in surface-brightness behavior seen
across the Voorwerp (section 5.2.1). It may therefore be more likely that the ionization pattern records
a complex luminosity history of the AGN.
	
\subsection{Embedded star formation\label{starform}}
The ionization level and temperature of the gas in Hanny's Voorwerp require the dominant ionization
mechanism to be photoionization by a continuum extending well into the far-UV. However, we find 
evidence of star formation in a few isolated regions both from continuum and emission-line properties.

In all three broad bands (F225W, F814W, F160W) there are several continuum sources embedded in the small bright northern part of the Voorwerp, nearest IC 2497. Each is associated with an area of very low [O III]/H$\alpha$ (although there are other nearby
regions of similar line ratio without an obvious embedded continuum object); both lines
show local maxima at their locations. The brightest of these are detected in the mid-UV image.
These are spatially resolved in images including line emission, but correcting the F814W image for this (using the H$\alpha$ image as a template for the Paschen continuum and weak emission lines) shows the brightest to be unresolved (Fig. \ref{fig-starknots}). This nominally pure-continuum image suggests that additional
continuum objects are present within this single 2" region at lower signal-to-noise. The resolution
limit is roughly 2 F814W pixels in WFC3, or 0.08" (80 pc). The associated H$\alpha$ emission regions
have been investigated using the surrounding [O III]/H$\alpha$ values to correct the H$\alpha$
image for AGN-ionized gas. These H$\alpha$ regions have FWHM=4.3--4.9 pixels, or 3.8--4.5
pixels (150-180 pc) after making a Gaussian correction for instrumental resolution.

\begin{figure*}
\includegraphics[scale=0.7,angle=0]{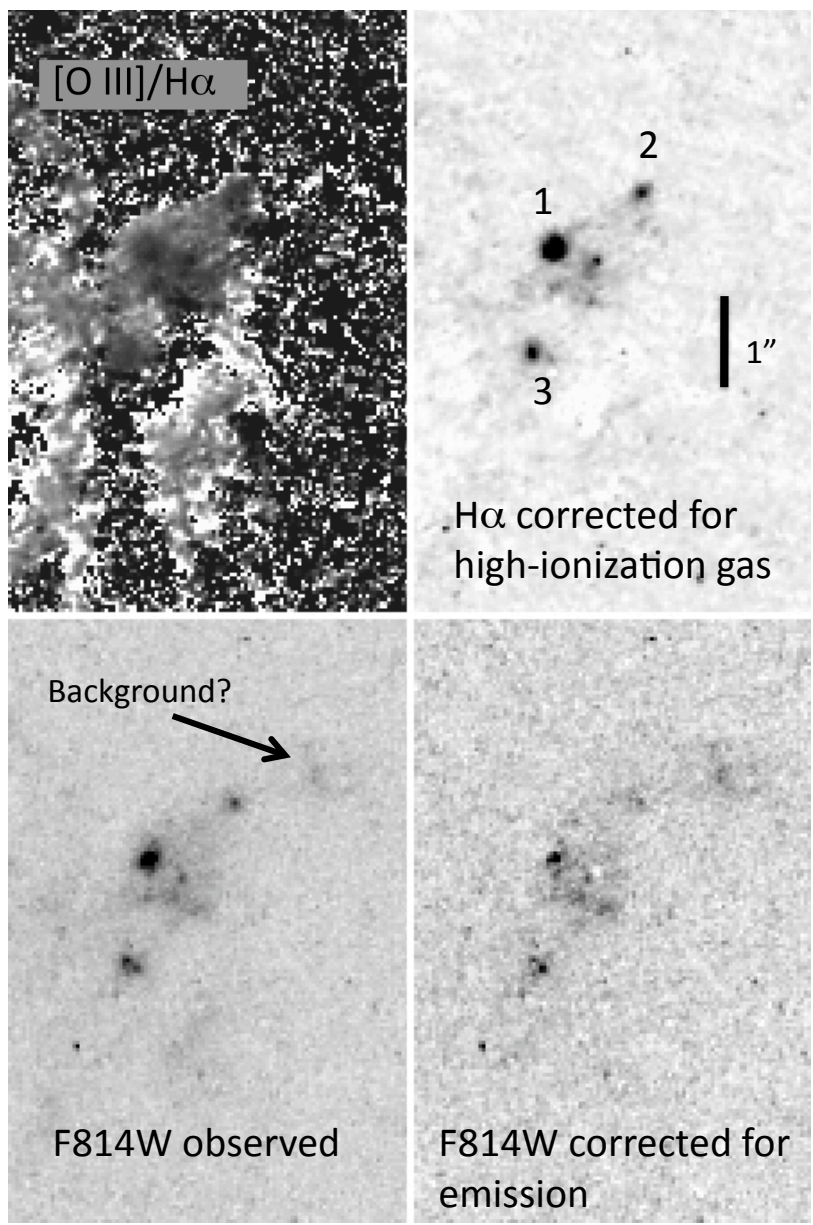}
\caption{The region containing candidate star-forming knots in the brightest part of Hanny's Voorwerp.  Upper left, the [O III]/H$\alpha$ ratio (dark is low, with strong H$\alpha$). Upper right, net H$\alpha$ from the low-ionization component, after subtracting a scaled version of the [O IIII] ACS image to reduce the contribution from the highly-ionized component. Lower left, the observed F814W image. Lower right, F814W after approximate correction for gaseous emission in lines and the Paschen continuum, using a scaled version of the H$\alpha$ image. This should be the most accurate rendition of the starlight continuum. The candidate star-forming knots are unresolved in the continuum, but resolved at the 0.2" level in H$\alpha$. The three brightest regions listed in Table \ref{tbl-sfregions} are marked, along with a much redder object which may be a background galaxy.}
 \label{fig-starknots}
\end{figure*}

Table \ref{tbl-sfregions} presents aperture photometry (within a matched radius of 0.38") for the
three brightest regions, noting that they exhibit various degrees of substructure down to the
resolution limit. The colors are very similar, especially $I_{814}-H_{160}$ where the errors are small.
They also lie close to the continuum shape for a young stellar population. Fig. \ref{fig-starform}
compares their broadband spectral energy distributions (SEDs) to the predictions of the
Starburst99 models \citep{starburst99}. The continuum slope we observe allows virtually any age (for either a short 
burst or ongoing star formation) up to $\approx 2 \times 10^8$ years. The H$\alpha$ emission associated with these star-forming
regions shows that star formation continues at a significant level, favoring either smaller ages or extended
star-foring timescales. This uncertainty translates into very poor constraints on the total stellar mass formed.

\begin{figure*}
\includegraphics[scale=0.5,angle=90]{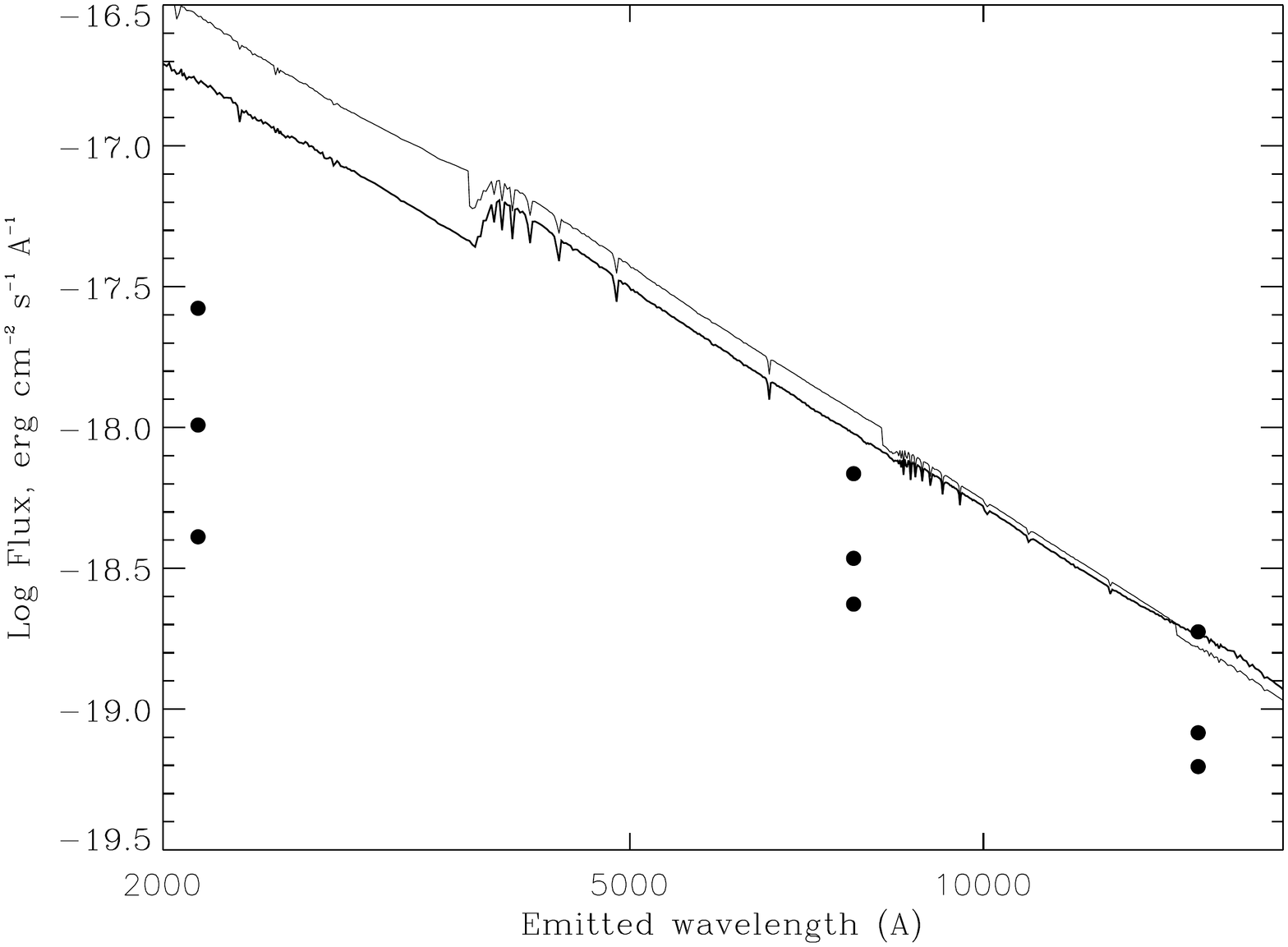}
\caption{Spectral energy distributions (SEDs) of star-forming knots in Hanny's Voorwerp
from continuum images, compared to predicted spectra of aging star-forming regions
from the Starburst99 models. 
These models have Z=0.004 to roughly match the spectroscopic gas abundances in the object. 
The SEDs are broadly consistent with clusters at a range of ages and reddening; the presence of large H II regions around
them, and lack of a cutoff at least to wavelengths as short as the F225W UV points, favors young clusters or assemblages with ongoing star formation,
with effective extinction as large as 2.5 magnitudes at 2250 \AA\  ($A_v \approx 1.0 $). }
\label{fig-starform}
\end{figure*}

We see no comparable blue objects elsewhere in the field outside IC 2497; no luminous
star-forming regions or bright stellar clusters of greater age appear in the Voorwerp or in the 
region we cover of the H I tail. The star formation is uniquely associated with a small region,
at least approximately aligned with the small-scale jet and larger outflow seen in radio observations
as well as the emission-line ``fingers".

We note a very red object at the western end of this brightest region, briefly mentioned by
\cite{Lintott09}. Unlike the other continuum structures, it is not associated with an emission-line peak
(or any detected emission-line structure). Several additional objects with similar extent and
color are seen in both F814W and F160W, possibly background galaxies.

In contrast to
Minkowski's Object, the overall ionization here is maintained by an AGN rather than the
local stellar populations; only with the spatial resolution of HST have we been able to distinguish their
effects. The star formation, although less dominant energetically, may have shared the same
general origin; as discussed by \cite{Fragile2004} and \cite{Croft2006} for Minkowski's Object,
the external pressure of a radio jet can compress pre-existing material, even if hot and at low density,
strongly enough to drive cooling, fragmentation, and presumably star formation. The H I cloud
around IC 2497 would provide a rich target for such a process, cooler than envisioned 
in their calculations and thus needing less compression to begin collapse. In having such a
pre-existing reservoir of cool H I, the situation here may be more like that in Centaurus A,
where the jet impinges on dense clouds. The distinction has been stressed by \cite{Croft2006},
noting different morphological relations between the young stars and H I in Cen A and
Minkowski's Object. 

The level of star formation we find in Hanny's Voorwerp is modest both in comparison to IC 2497 and to such 
environments as Minkowski's Object. Making an approximate allowance for fainter H$\alpha$ regions beyond the apertures in table \ref{tbl-sfregions} adding as much as an additional 30\%, the H$\alpha$
luminosity of these star-forming regions is in the range 0.9-1.4$\times 10^{40}$ erg s$^{-1}$,
or 14--21\% of the value in Minkowski's Object \citep{Croft2006}. We do note that additional star formation, if not collected into clusters, could be hidden
in the diffuse UV emission from the gas; the clusters account for only a few per cent of the integrated flux found by Swift and GALEX,
which is predominantly continuum. This continuum certainly has a substantial contribution in this band from nebular processes, and
might also include scattered AGN light from embedded dust.


\section{Summary}

Our data support the scheme of a quasar rapidly dropping in luminosity; STIS spectra show no highly-ionized gas in IC 2497. However, the STIS
spectra and images show signs of outflow from the nucleus, leading to the possibility that some of the energy output from the AGN has
switched from radiation to kinetic form. This is seen in an expanding loop of gas $\approx 500 $ pc in diameter, and in star-forming
regions within Hanny's Voorwerp which we attribute to compression by the outflow seen at radio wavelengths.

We suggest the following sequence of events: A major merger liberated the massive tail of H I, with geometry such that the remnant IC 2497 retained
its disk although significantly warping it. The low metallicity of the gas suggests that it began in the extreme outer disk of the galaxy.
The dynamical disturbance eventually triggered an episode of accretion into the central black hole, with
an ionizing luminosity appropriate for a quasar. The escaping UV radiation ionized parts of the H I tail, creating Hanny's Voorwerp. Recently (perhaps a
few million years before our current view), an outflow from the core began, including the small-scale radio jet, the emission-line ring near the
nucleus, and a narrow outflow directed roughly toward Hanny's Voorwerp, triggering star formation in the small region where this outflow impinged on
relatively dense gas. Then within the last 100,000 years before our view of the galaxy nucleus, the ionizing luminosity dropped enormously,
by 2--4 orders of magnitude, leaving Hanny's Voorwerp as the only remaining evidence of this episode. The close association of this drop in time with 
the onset of outflows may indicate that this was not completely a corresponding drop in the level of activity of the nucleus, but rather a switch between
so-called quasar and radio modes of accretion. The light-travel delay across the system is small compared to limits on the age of the outflows, so that
we cannot say whether this fading was a one-time event or recurs after multiple bright episodes.
 
The massive H I tail is crucial in making the history of the AGN observable. Since such tails could be regarded as ``living fossils" of epochs when major mergers
were common, we might expect to see more objects of this kind as data of the necessary spatial resolution become available at larger redshifts. If the
combination of fading ionization and onset of outflows proves to be common, we might also find isolated star-forming regions near formerly
active galaxies along the axis of an outflow, outlasting all but the very low-frequency radio emission associated with the outflow itself.

This nearest quasar escaped all the standard optical, X-ray, and radio survey techniques. The fact that  IC 2497 is at such a low redshift
(and would have been the nearest AGN of its luminosity) would be very unlikely unless similar episodic behavior is common among
AGN. Indeed, a dedicated search by Galaxy Zoo participants has found additional, less luminous examples of large-scale
ionized clouds, including additional potential examples of fading AGN, in the local Universe \citep{Keel11}. These are
found systematically in disturbed systems, attesting to the importance of an extended reservoir of neutral gas as a probe of the pattern and history of
ionizing radiation from AGN. Additional surveys are in progress, keying on [O III] emission as a sensitive tracer of highly-ionized gas
around samples of galaxies with and without detected AGN.

\acknowledgments
We thank Ski Antonucci, Jay Anderson, and Max Mutchler for useful exchanges on the handling and
interpretation of these data. Ray White, Brian Skiff, and Gary Ferland made helpful comments during the
preparation of this paper.

Support for program number 11620 was provided by NASA through a grant from the Space Telescope Science Institute, which is operated by the Association of Universities for Research in Astronomy, Incorporated, under NASA contract NAS5-26555. This work was based in part on observations made with the 
NASA Galaxy Evolution Explorer. GALEX was operated for NASA by the California Institute of Technology under NASA contract NAS5-98034.
Support for the work of K.S. was provided by NASA through Einstein Postdoctoral Fellowship grant number PF9-00069 issued by the Chandra X-ray Observatory Center, which is operated by the Smithsonian Astrophysical Observatory for and on behalf of NASA under contract NAS8-03060.

W.C. Keel acknowledges support from a Dean's Leadership Board faculty fellowship.
Galaxy Zoo was made possible by funding from a Jim Gray Research Fund from Microsoft, and The Leverhulme Trust. S. D. Chojnowski participated through the SARA
Research Experiences for Undergraduates program funded by the US National Science
Foundation. Funding for the creation and distribution of the SDSS Archive has been
provided by the Alfred P. Sloan Foundation, the Participating Institutions,
the National Aeronautics and Space Administration, the National Science
Foundation, the U.S. Department of Energy, the Japanese Monbukagakusho,
and the Max Planck Society. The SDSS Web site is http://www.sdss.org/. 
The SDSS is managed by the Astrophysical Research Consortium (ARC) for
the Participating Institutions. The Participating Institutions are The
University of Chicago, Fermilab, the Institute for Advanced Study, the Japan
Participation Group, The Johns Hopkins University, Los Alamos National
Laboratory, the Max-Planck-Institute for Astronomy (MPIA), the
Max-Planck-Institute for Astrophysics (MPA), New Mexico State
University, Princeton University, the United States Naval Observatory, and
the University of Washington. STSDAS and PyRAF are products of the Space 
Telescope Science Institute, which is operated by AURA for NASA.
This publication makes use of data products from the Wide-field Infrared 
Survey Explorer, which is a joint project of the University of California, Los Angeles, 
and the Jet Propulsion Laboratory/California Institute of Technology, funded by the 
National Aeronautics and Space Administration.




\clearpage







\clearpage


\clearpage

\begin{table}
\begin{center}
\caption{New data\label{tbl-1}}
\begin{tabular}{lcccr}
\tableline\tableline
Instrument & Mode  &  Spectral element  & ObsID	& Exposure, s \\
\tableline
HST	& WFC3 & F225W	 & ib5603020 & 2952\\
	& &	F814W	 & ib5603030 & 2952\\
	& &      F160W	& ib5603010 & 2698\\
	& ACS& 	FR505N\#5240	& jb5602010 & 2570\\
	& &	FR716N\#6921& jb5602020& 2750\\
	& STIS	& G750L	 & ob5601010 & 2000\\
	& & 	  G430L	 & ob5601030 & 2750\\
GALEX & image & NUV & 	07061-IC2497	& 6193\\
	& spectrum & NUV  &		07061-IC2497	& 17798\\
	& image & FUV  &	07061-IC2497	& 6193\\
	& spectrum & FUV &	07061-IC2497&	17798\\
WIYN & OPTIC	& V	& -- & 1200\\
	& & 	B & -- & 	1200\\
	& & 	I & -- & 		600\\
KPNO 2.1m & GoldCam & 26new 600 l  mm$^{-1}$ & -- & 2700 \\ 
\tableline
\end{tabular}

\end{center}
\end{table}
\clearpage

\begin{table}
\begin{center}
\caption{Integrated UV/Optical Line Fluxes from Hanny's Voorwerp\label{tbl-galex}}
\begin{tabular}{lc}
\tableline\tableline
Line   & Flux ($10^{-15}$ erg cm$^{-2}$ s$^{-1}$) \\
\tableline
C IV $\lambda 1549$ & 51 \\
He II $\lambda 1640$ & 50 \\
C III] $\lambda 1909$ & 49 \\
{} [Ne IV] $\lambda 2425$ & 10\\
He II $\lambda 4686$ & 6.7 \\
{} [O II] $\lambda 3727$ &  27 \\ 
{} [O III] $\lambda 5007$&  191 \\
H$\alpha$ & 62 \\
\tableline
\end{tabular}
\clearpage

\end{center}
\end{table}

\begin{table}
\begin{center}
\caption{Nuclear emission-line structures\label{tbl-nuclei}}
\begin{tabular}{lcc}
\tableline\tableline
Region        &    Nucleus         & Loop \\
\tableline
$z$              &    0.04987       & 0.05142  \\
Balmer FWHM (km s$^{-1}$)  & 1060   & 790 \\
\ [O III] FWHM (km s$^{-1}$) &  620    & 800      \\
\ [O III]/H$\beta$  &       0.67    &     4.1 \\
\ [N II]/H$\alpha$ &      2.17     &    2.14 \\
\  [S II]/H$\alpha$ &     0.60   & 0.86  \\
\ [O II] $\lambda 3727$/[O III] $\lambda 5007$ & 2.29 & 1.20\\
H$\alpha$/H$\beta$ &  4.35  &   3.9  \\
\tableline
\end{tabular}

\end{center}
\end{table}
\clearpage

\begin{table}
\begin{center}
\caption{Properties of candidate star-forming regions\label{tbl-sfregions}}
\begin{tabular}{lccc}
\tableline\tableline
Region   & 1 & 2 & 3 \\
\tableline
$\alpha$ (2000) &   9:41:03.682  &  9:41:03.607  &  9:41:03.703  \\
$\delta$  (2000) &    +34:43:41.42 &  +34:43:42.04 &  +34:43:40.28 \\
F225W flux ($10^{-20}$ erg cm$^{-2}$ s$^{-1}$ \AA $^{-1}$) & 265 & 102 & 40.9 \\
F814W flux ($10^{-20}$ erg cm$^{-2}$ s$^{-1}$ \AA $^{-1}$) & 68.5 & 23.6 & 34.3 \\
F160W flux ($10^{-20}$ erg cm$^{-2}$ s$^{-1}$ \AA $^{-1}$) & 18.8 &6.3 & 8.2 \\
H$\alpha$ flux ($10^{-16}$ erg cm$^{-2}$ s$^{-1}$) & 11.6 & 3.54 & 5.15 \\ 
\tableline
\end{tabular}

\end{center}
\end{table}


\begin{thebibliography}{}

\bibitem[Anderson 
\& Bedin(2010)]{CTE} Anderson, J., \& Bedin, L.~R.\ 2010, PASP 122, 1035


\bibitem[Bennert et 
al.(2006)]{Bennert} Bennert, N., Jungwiert, B., Komossa, S., Haas, M., \& Chini, R.\ 2006, \aap, 446, 919 

\bibitem[Binette 
\& Robinson(1987)]{Binette1987} Binette, L., \& Robinson, A.\ 1987, \aap, 177, 11 

\bibitem[Boquien et 
al.(2007)]{Boquien} Boquien, M., Duc, P.-A., Braine, J., Brinks, E., Lisenfeld, U., \& Charmandaris, V.\ 2007, \aap, 467, 93 

\bibitem[Briggs et 
al.(2001)]{Briggs} Briggs, F.~H., M{\"o}ller, O., Higdon, J.~L., Trentham, N., \& Ramirez-Ruiz, E.\ 2001, \aap, 380, 418 


\bibitem[Churazov  et al.(2005)]{Churazov} Churazov, E., Sazonov,   
S., Sunyaev, R., Forman, W., Jones, C., \& Bo\"hringer, H.\ 2005, \mnras, 363, L91 

\bibitem[Crenshaw et al.(2010)]{Crenshaw2010} Crenshaw, D.~M., 
Kraemer, S.~B., Schmitt, H.~R., Jaff{\'e}, Y.~L., Deo, R.~P., Collins, 
N.~R., \& Fischer, T.~C.\ 2010, \aj, 139, 871 

\bibitem[Croft et al.(2006)]{Croft2006} Croft, S., et al.\ 2006, 
\apj, 647, 1040 


\bibitem[Done 
\& Gierli{\'n}ski(2005)]{Done} Done, C., \& Gierli{\'n}ski, M.\ 2005, \mnras, 364, 208 

\bibitem[Evans et al.(2008)]{Evans} Evans, D.~A., et al.\ 
2008, \apj, 675, 1057 

\bibitem[Fabian et al.(2008)]{Fabian2008} Fabian, A.~C., 
Johnstone, R.~M., Sanders, J.~S., Conselice, C.~J., Crawford, C.~S., 
Gallagher, J.~S., III, \& Zweibel, E.\ 2008, \nat, 454, 968 


\bibitem[Fragile et al.(2004)]{Fragile2004} Fragile, P.~C., Murray, 
S.~D., Anninos, P., \& van Breugel, W.\ 2004, \apj, 604, 74 

\bibitem[Fu \& Stockton(2007)]{FuStockton07} Fu, H. \& Stockton, A. 2007, 
ApJ 666, 794

\bibitem[Fu \& Stockton(2009)]{FuStockton} Fu, H. \& Stockton, A. 2009, 
ApJ 690, 593

\bibitem[Gil de Paz et al.(2007)]{NGS} Gil de Paz, A., et 
al.\ 2007, \apjs, 173, 185 

\bibitem[Goodman(2003)]{Goodman} Goodman, J.\ 2003, \mnras, 
339, 937 

\bibitem[Graham 
\& Price(1981)]{GrahamPrice} Graham, J.~A., \& Price, R.~M.\ 1981, \apj, 247, 813 

\bibitem[Groves et al.(2004)]{Groves04} Groves, B.A., Dopita, M.A, \& Sutherland, R.S. 2004 ApJS 153, 9

\bibitem[Groves et al.(2006)]{Groves06} Groves, B.A., Heckman, T.M., \& Kauffmann, G. 2006 MNRAS 371, 1559

\bibitem[Harris et al.(2010)]{Harris2010} Harris, G.~L.~H., 
Rejkuba, M., \& Harris, W.~E.\ 2010, \pasa, 27, 457 

\bibitem[Hopkins et al.(2005)]{Hopkins2005} Hopkins, P.~F., 
Hernquist, L., Martini, P., et al.\ 2005, \apjl, 625, L71 

\bibitem[Hota et al.(2011)]{Hota} Hota, A., Sirothia, S.~K., 
Ohyama, Y., et al.\ 2011, \mnras, 417, L36 

\bibitem[Hibbard et al.(2001)]{Rogues} Hibbard, J.~E., van 
Gorkom, J.~H., Rupen, M.~P., 
\& Schiminovich, D.\ 2001, Gas and Galaxy Evolution, 240, 657 

\bibitem[Howard et al.(1993)]{simatlas} Howard, S., Keel, W.~C., 
Byrd, G., \& Burkey, J.\ 1993, \apj, 417, 502 

\bibitem[Janiuk et al.(2004)]{Janiuk} Janiuk, A., Czerny, B., 
Siemiginowska, A., \& Szczerba, R.\ 2004, \apj, 602, 595 

\bibitem[Jarett et al. (2011)]{Jarrett}Jarett, T.H. et al. 2011, ApJ 735, 112

\bibitem[J{\'o}zsa et 
al.(2009)]{Josza} J{\'o}zsa, G.~I.~G., et al.\ 2009, \aap, 500, L33 


\bibitem[Keel(1989)]{KeelCenA} Keel, W.~C.\ 1989, European 
Southern Observatory Conference and Workshop Proceedings, 32, 427 

\bibitem[Keel et al.(2006)]{Keel0313} Keel, W.~C., White, R.~E., 
III, Owen, F.~N., \& Ledlow, M.~J.\ 2006, \aj, 132, 2233 

\bibitem[Keel et al.(2012)]{Keel11} Keel, W.C. et al. 2012, MNRAS, 420, 878

\bibitem[Kirkman 
\& Tytler(2008)]{Kirkman} Kirkman, D., \& Tytler, D.\ 2008, \mnras, 391, 1457 

\bibitem[Koekemoer et al.(2002)]{mdriz} Koekemoer, A.~M., 
Fruchter, A.~S., Hook, R.~N., 
\& Hack, W.\ 2002, The 2002 HST Calibration Workshop : Hubble after the Installation of the ACS and the NICMOS Cooling System, 337 

\bibitem[Leitherer et al.(1999)]{starburst99} Leitherer, C., et 
al.\ 1999, \apjs, 123, 3 

\bibitem[Lintott et al.(2008)]{Lintott08} Lintott, C.~J., et al.\ 
2008, \mnras, 389, 1179 

\bibitem[Lintott et al.(2009)]{Lintott09} Lintott, C.~J., et al.\ 
2009, \mnras, 399, 129 

\bibitem[Lupton et al.(2004)]{2004PASP..116..133L} Lupton, R., Blanton, 
M.~R., Fekete, G., Hogg, D.~W., O'Mullane, W., Szalay, A., 
\& Wherry, N.\ 2004, \pasp, 116, 133 

\bibitem[MacAlpine et al.(1985)]{MacAlpine} MacAlpine, G., Davidson, K., Gull, T.R.,
\& Wu, C.-C. 1985, ApJ 294, 147

\bibitem[Maccarone et al.(2003)]{Maccarone} Maccarone, T.~J., 
Gallo, E., \& Fender, R.\ 2003, \mnras, 345, L19 

\bibitem[Malkan et al.(1998)]{Malkan} Malkan, M.~A., Gorjian, 
V., \& Tam, R.\ 1998, \apjs, 117, 25 

\bibitem[Malphrus et al.(1997)]{Malphrus} Malphrus, B.~K., 
Simpson, C.~E., Gottesman, S.~T., \& Hawarden, T.~G.\ 1997, \aj, 114, 1427 

\bibitem[Martini 
\& Schneider (2003)]{Martini2003} Martini, P., \& Schneider, D.~P.\ 2003, \apjl, 597, L109 

\bibitem[Martini(2004)]{Martini2004} Martini, P.\ 2004, Coevolution 
of Black Holes and Galaxies, 169 

\bibitem[McHardy et al.(2006)]{McHardy} McHardy, I.~M., 
Koerding, E., Knigge, C., Uttley, P., 
\& Fender, R.~P.\ 2006, \nat, 444, 730 

\bibitem[Merloni \& Heinz (2007)]{MH2007}  Merloni, A. \& Heinz, S. 2007 \mnras 381, 589     

\bibitem[Merloni \& Heinz (2008)]{MH2008}  Merloni, A. \& Heinz, S.  2008, \mnras, 388, 1011   

\bibitem[Michel-Dansac et al.(2010)]{Leo2010} Michel-Dansac, 
L., et al.\ 2010, \apjl, 717, L143 

\bibitem[Morganti et al.(1992)]{morganti92} Morganti, R., Fosbury, 
R.~A.~E., Hook, R.~N., Robinson, A., 
\& Tsvetanov, Z.\ 1992, \mnras, 256, 1P 

\bibitem[Morrissey et al.(2007)]{GALEXdata} Morrissey, P., et 
al.\ 2007, \apjs, 173, 682 

\bibitem[Netzer(1990)]{Netzer90} Netzer, H.\ 1990, in Active Galactic Nuclei (eds. R.D. Blandford, H. Netzer, \& L. Woltjer), 57 (Springer-Verlag) 

\bibitem[Osterbrock \& Ferland (2006)]{AGNAGN} Osterbrock, D.E. \& Ferland, G.J., 2006,  {\it Astrophysics of Gaseous Nebulae and Active
Galactic Nuclei}, 2nd ed., University Science Books (Sausalito, California)


\bibitem[Pier et al.(2003)]{Pier} Pier, J.~R., Munn, J.~A., 
Hindsley, R.~B., Hennessy, G.~S., Kent, S.~M., Lupton, R.~H., 
\& Ivezi{\'c}, {\v Z}.\ 2003, \aj, 125, 1559 

\bibitem[Rampadarath et al.(2010)]{Rampadarath} Rampadarath, H. et al. 2010, A\&A 517, L8

\bibitem[Schawinski et al.(2010)]{Schawinski2010}Schawinski, K. et al. 2010, \apj 724, L30

\bibitem[Schegel et al.(1998)]{Schlegel} Schlegel, D.J., Finkbeiner, D.P.  \& Davis, M. 198, ApJ 500, 525

\bibitem[Schneider et al.(1989)]{Leo1989} Schneider, S.~E., et 
al.\ 1989, \aj, 97, 666 

\bibitem[Shields 
\& Wheeler(1978)]{Shields1978} Shields, G.~A., \& Wheeler, J.~C.\ 1978, \apj, 222, 667 



\bibitem[Stern et al.(2012)]{Stern} Stern, D. et al. 2012,  ApJ in press (arXiv:1205.0811)

\bibitem[Su et al.(2010)]{Fermi} Su, M., Slatyer, T.~R., 
\& Finkbeiner, D.~P.\ 2010, \apj, 724, 1044 

\bibitem[Sutherland et al.(1993)]{CenAshocks} Sutherland, R.~S., 
Bicknell, G.~V., \& Dopita, M.~A.\ 1993, \apj, 414, 510 

\bibitem[Tonry et al.(1997)]{OTCCD} Tonry, J., Burke, B.~E., 
\& Schechter, P.~L.\ 1997, \pasp, 109, 1154 

\bibitem[Wright et al. (2010)]{WISE} Wright, E.L.
et al. 2010, AJ 140, 1868



\end{thebibliography}
\end{document}